\documentclass[12pt, a4paper]{article} 

\usepackage{graphicx}
\usepackage{url}

\usepackage{listings}
\usepackage{xcolor}

\definecolor{codegreen}{rgb}{0,0.6,0}
\definecolor{codegray}{rgb}{0.5,0.5,0.5}
\definecolor{codepurple}{rgb}{0.58,0,0.82}
\definecolor{backcolour}{rgb}{0.95,0.95,0.92}

\lstdefinestyle{mystyle}{
    backgroundcolor=\color{backcolour},   
    commentstyle=\color{codegreen},
    keywordstyle=\color{magenta},
    numberstyle=\tiny\color{codegray},
    stringstyle=\color{codepurple},
    basicstyle=\ttfamily\footnotesize,
    breakatwhitespace=false,         
    breaklines=true,                 
    captionpos=b,                    
    keepspaces=true,                 
    numbers=left,                    
    numbersep=5pt,                  
    showspaces=false,                
    showstringspaces=false,
    showtabs=false,                  
    tabsize=2
}

\usepackage{comment} 
\usepackage[utf8]{inputenc} 
\usepackage[T1]{fontenc} 
\usepackage{amssymb}
\usepackage{booktabs}

\usepackage{amsmath,amsfonts,amsthm} 
\usepackage{tikz} 
\usepackage{tikz-cd}
\usetikzlibrary{positioning,arrows} 
\usetikzlibrary{decorations.pathreplacing} 
\usepackage{tikzsymbols} 
\usepackage{blindtext} 

\usepackage{geometry} 

\usepackage{setspace} 
\usepackage{subcaption}
\usepackage[T1]{fontenc} 
\usepackage[utf8]{inputenc} 

\usepackage{fancyhdr} 
\usepackage{gensymb}
\usepackage{multirow}

\usepackage[
    backend=bibtex, 
    natbib=true,
    style=phys,
    biblabel=brackets,
    giveninits=true,
    abbreviate=false,
    doi=false, url=false, isbn=false,
    block=space,
]{biblatex}
\title{Bayesian Analysis of Conventional and Ultrafast Spectroscopy Data for Investigating Detachment in the MAST-Upgrade Super-X} 
\author{
Xander Pope \\
\textbf{Project Supervisors:} Prof. B, Lipschultz, Dr. K, Verhaegh,\\ Dr. C, Bowman}

\date{} 

\begin{document}



\begin{titlepage}

	
\begin{minipage}{0.4\textwidth} 
    \begin{flushleft} 
    \large
    University of York\\ 
    School of Physics, Engineering and Technology\\ 
    2022/23\\ 
    Fusion Energy\\ 
    Supervisors:\\ Prof. B, Lipschultz, \\Dr. K, Verhaegh, \\Dr. C, Bowman 
    \end{flushleft}
\end{minipage}
	
\vspace*{2in} 
	
\center 

	
{\huge\bfseries BAYESIAN ANALYSIS OF CONVENTIONAL AND ULTRAFAST SPECTROSCOPY DATA FOR INVESTIGATING DETACHMENT IN THE MAST-UPGRADE SUPER-X}\\[0.4cm] 
XANDER POPE 
	
\vfill 


\vfill 


\vfill 


\vfill 
	

\end{titlepage}

\maketitle 
\begin{center} 
\end{center} 


\begin{abstract}

This paper presents the application, testing and first results of a new adaptive Bayesian inference analysis which utilises conventional and ultrafast spectroscopic measurements made in the divertor chamber to investigate the divertor physics during detachment.
Validation of this software is performed prior and during analyses of results, demonstrated by compelling reproductions of ideal test cases and synthetic spectroscopic measurements.
Application on real diagnostic data shows strong agreement with results from previous analysis methods.
We identify unprecedented success in significant advances in time and computational efficiencies.
We demonstrate a $\lesssim$1000$\times$ reduction in analysis time for spectroscopic measurements from simulated and real Super-X configurations, with the analysis technique presented in this report completing in <3 minutes.
Analysis of synthetic and real diagnostic measurements identifies detachment physics in agreement with previous literature.


\end{abstract}
\clearpage
\setcounter{page}{1} 

\section{Introduction}




Many challenges line the road to commercial fusion energy, a particularly potent obstacle is that of exhausting heat from the main chamber.

Fusing deuterium (D) and tritium (T) into $^4$He ($\alpha$-particle) forms the most viable fusion reaction (Eq. \ref{eq:DTReaction}).
Our concern is for the 20\% of the reaction energy that remains in the core plasma as $\alpha$-particles.
Radial diffusion causes this power to ultimately cross the last closed flux surface and enter the scrape-off-layer (SOL).
\begin{equation}
    \label{eq:DTReaction}
    D + T \rightarrow n (14.1\mathrm{MeV}) + \alpha (3.5\mathrm{MeV})
\end{equation}

Generally, the radial movement of plasma across flux surfaces is very slow compared to movement along flux surfaces.
The SOL travels parallel to flux surfaces.
Hence, the SOL spatially reduces the gradually diffused power into an intense heat flux, directed towards a solid target in the divertor.
Such flux would need to be significantly reduced to meet current material limits for the divertor \cite{2019Iter, 2014Demo}.
Commercial viability, in terms of exhausting residual heat, is hence hindered until this heat flux can be mitigated.

Detachment results in a simultaneous reduction of the target ion flux and electron temperature, which can reduce the target heat flux by orders of magnitude.
This is achieved by simultaneous power, momentum and particle (i.e., ion) losses as a result of plasma-atom/molecule interactions \cite{2019Atomic, 2021Molecular}.
To access detachment, the divertor region must exist at a temperature below $\approx$5eV \cite{2023Detachment}.
These conditions can be accomplished by a number of methods, including: inducing radiative losses in the plasma by seeding impurities, increasing the collisionality of the plasma by increasing the particle density in the core or divertor and investigating alternative divertor configurations (ADC).

\begin{figure}
    \centering
    \includegraphics[width=0.9\textwidth]{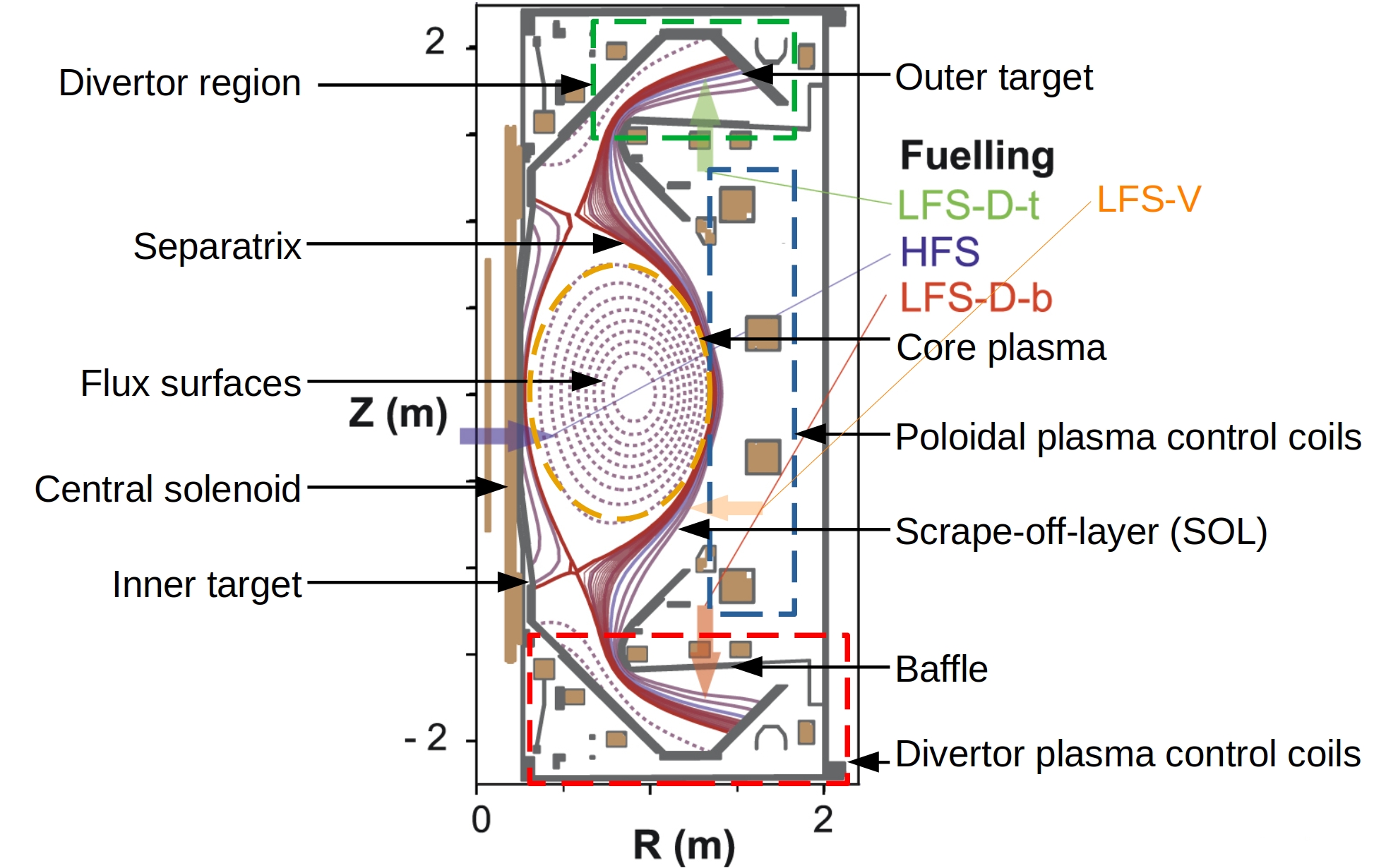}
    \caption{Cross-section of the MAST-U tokamak. The Super-X divertor is seen above and below the main chamber. Gas fuelling locations are shown at the high field and low field sites (HFS and LFS respectively). 
    LFS can be done within the top (t) and bottom (b) divertors (D), or within the vacuum vessel itself (V).
    Adapted from \cite{2022Spectroscopic}.}
    \label{fig:SuperX}
\end{figure}

ADCs, such as the Super-X divertor, developed for use on the MAST-U tokamak (Fig. \ref{fig:SuperX}), utilise magnetic shaping of the divertor topology to improve power exhaust and improve access to divertor detachment.
By shifting the outer target to a larger major radius, the Super-X increases the surface area of the target.
Furthermore, the plasma is moved towards a lower poloidal flux density which in turn causes the plasma to spread and the heat flux to lower.
The Super-X divertor is also installed on both the top and bottom of the tokamak allowing a second outer divertor.
Additionally, to ensure the confinement of neutral atoms/molecules that cannot be confined electromagnetically within the divertor region, the Super-X implements a baffle.
This facilitates further detachment-related interactions within the divertor.

\subsection{Detachment}

\subsubsection{Particle and Power Balance}

\begin{figure}
    \centering
    \includegraphics[width=0.7\textwidth]{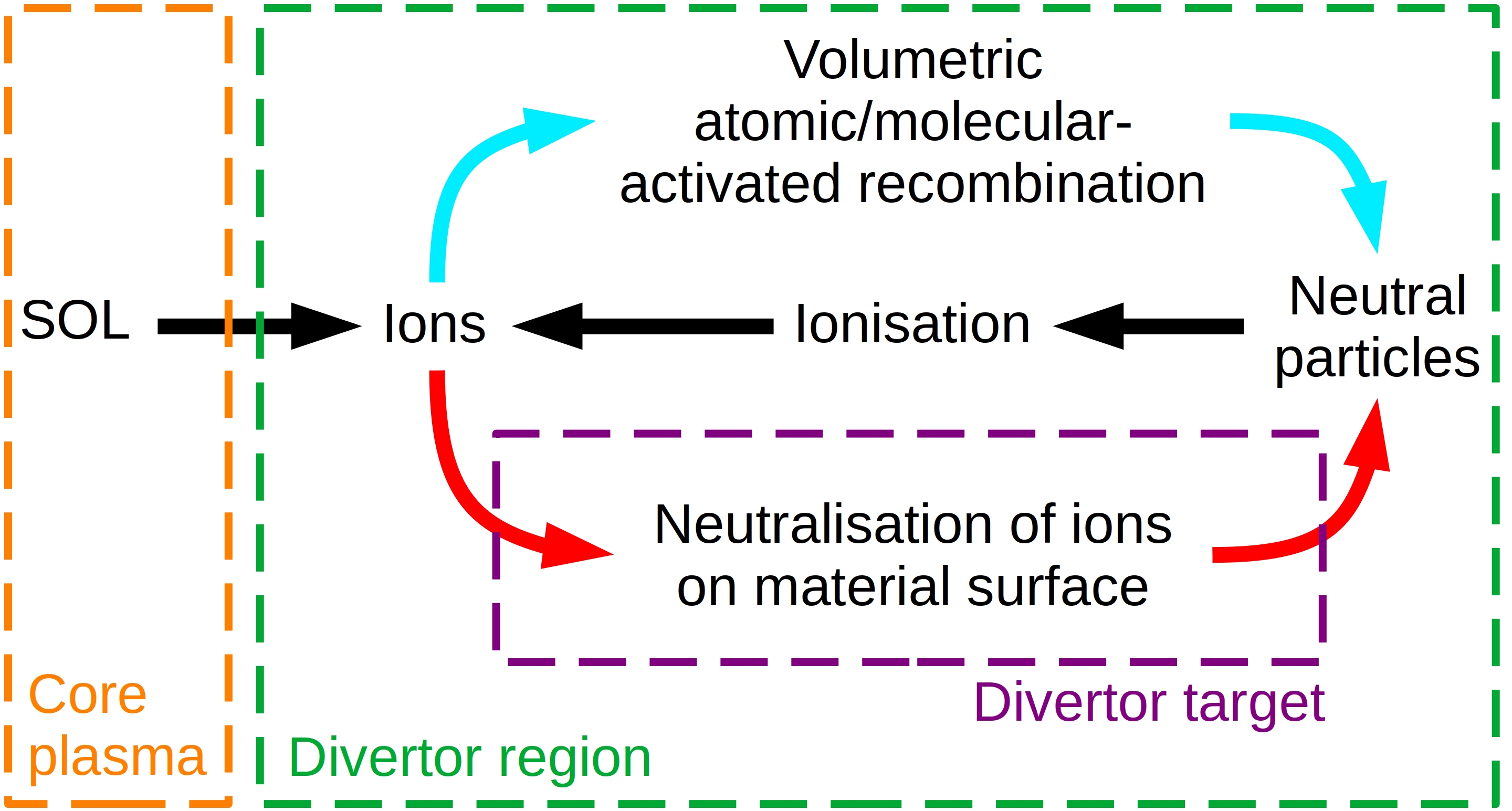}
    \caption{Summary of processes creating/destroying ions within the recylcing region of the divertor. Adapted from \cite{2017Recycling}.}
    \label{fig:DivertorRecycling}
\end{figure}

Plasma in the SOL enters a region of recycling conditions once in the divertor, shown in Fig. \ref{fig:DivertorRecycling}.
Within these conditions ions continuously neutralise/recombine and re-ionise.
It is the neutralisation on the target that transfers power and momentum, initially from the core plasma, onto the material.
When the bulk of this recycling is occurring on the target, the system is `attached'.
Thus the flux of ions reaching the target must be mitigated by `detaching' this high recycling region from the target and into the divertor chamber.

Total ion flux incident on the target ($I_t$) is promoted by ion sources within the divertor ($I_i$) and any net influx of ions from the SOL ($I_\mathrm{SOL}$) and mitigated by ion sinks within the divertor ($I_r$) (Eq. \ref{eq:BasicParticleBalance}) \cite{2023Detachment}.
\begin{equation}
    \label{eq:BasicParticleBalance}
    I_t = I_i - I_r + I_\mathrm{SOL}
\end{equation}

Shown in \cite{2021Molecular, 2023Detachment}, combining Eq. \ref{eq:BasicParticleBalance} with power balance results in Eq. \ref{eq:UpdatedParticleBalance}, assuming $I_t \gg I_\mathrm{SOL}$.
$P_\mathrm{recl}$ is the power entering the recycling region, $E_\mathrm{ion}$ is the energy required to ionise hydrogen (inclusive of radiative losses due to preliminary excitation of the neutral before ionisation), $\gamma$ is the sheath transmission factor and $T_t$ the target temperature.
\begin{equation}
    \label{eq:UpdatedParticleBalance}
    I_t = \left(\frac{P_\mathrm{recl}}{E_\mathrm{ion}} - I_r\right) \times \frac{1}{1 + \frac{\gamma T_t}{E_\mathrm{ion}}}
\end{equation}

During deep detachment $\frac{\gamma T_t}{E_\mathrm{ion}} \ll 1$ \cite{2021Molecular}, we can approximate Eq. 3 into Eq. 4, giving further insight into what drives the ion target flux.
\begin{equation}
    \label{eq:FinalParticleBalance}
    I_t = \frac{P_\mathrm{recl}}{E_\mathrm{ion}} - I_r
\end{equation}

`Starving' the ion flux to the target \cite{2017Recycling, 1998PowerParticle} can be done by seeding large amounts of heavy impurities into the main chamber to facilitate power loss in the core plasma via radiation, resulting in reduced $P_\mathrm{recl}$ \cite{2007Seeding}.
This is not optimal for keeping the core plasma in fusion conditions, therefore, we must look to increase the ion sinks in the divertor to facilitate the detachment of the high recycling region.
It is important to note that a parallel line of understanding for facilitating detachment finds that pressure reductions at the divertor target (rather than power/particle losses), caused by volumetric momentum losses, are required \cite{2018Pressure}.
Whilst differing in approach, it is shown that both concepts are valid and equivalent in the requirement for simultaneous power, momentum and particle loss \cite{2019Compare}.

\subsubsection{Plasma-Atom and Plasma-Molecule Interactions}

A large number of plasma-atom and plasma-molecule interactions, occurring within the divertor region, provide the microscopic origin for detachment.
MAST-U is not equipped for use with tritium, henceforth interactions are discussed in terms of deuterium with note that these interactions apply for all hydrogen isotopes.

Interactions between the plasma and atoms/molecules in the divertor can directly or indirectly contribute to the ion flux reaching the target.
Interactions that provide ion sources (ionisation) and sinks (recombination) directly influence the balance of ions reaching the target, whilst interactions that produce a source of neutral atoms (dissociation) provide an indirect impact by inducing increased momentum losses through more plasma-neutral collisions.

Initially, the ion sources/sinks and power losses were studied spectroscopically through plasma-atom interactions (ionisation and electron-ion recombination (EIR)) \cite{2019Atomic, 2019Compare}. 
However, this could not explain the observed increase in Balmer-$\alpha$ emission during detachment. 
This required accounting for excited hydrogen atoms after molecular break-up when molecular ions ($D_2^+$, $D_2^-$ $\rightarrow$ $D^-$ + $D$) interact with the plasma \cite{2021Molecular}.

\begin{table}[]
\centering
\begin{minipage}[]{0.48\linewidth}
  \centering
  \begin{tabular}{cc}
    \hline
    Creation    &                                           \\ \hline
    1           & $D_2 + D^+ \rightarrow D_2^+ + D$           \\
    2           & $D_2 + e^- \rightarrow D_2^+ + 2e^-$      \\ \hline
    Destruction &                                           \\ \hline
    1           & $D_2^+ + e^- \rightarrow D + D^*$         \\
    2           & $D_2^+ + e^- \rightarrow D^+ + e^- + D^*$ \\
    3           & $D_2^+ + e^- \rightarrow 2D^+ + 2e^-$    
  \end{tabular}
  \caption{Plasma-molecule interactions involving $D_2^+$.
  Excited atoms denoted by $D^*$.}
  \label{tab:MolReactions}
\end{minipage}%
\begin{minipage}[]{0.48\linewidth}
  \centering
  \begin{tabular}{c|ccc}
     & D1  & D2  & D3     \\ \hline
    C1 & MAR & MAD & MAI    \\
    C2 & MAD & MAI & MAI x2
  \end{tabular}
  \caption{Combinations of creation (C) and destruction (D) events lead to different overall plasma-molecule interactions.}
  \label{tab:MolCombinations}
\end{minipage}
\end{table}

Whilst less understood than their atomic counterparts, plasma-molecule interactions (PMI) are known to be split into two stages: the creation of molecular ions ($D_2^+$ and/or $D_2^-$) followed by the destruction of these ions (Tab. \ref{tab:MolReactions}).
Only $D_2^+$ is considered in Tab. \ref{tab:MolReactions} as it is likely the dominant molecular ion \cite{2021Molecular}.
The combination of creation and destruction events determines whether there is a net; increase in atomic ions (Molecular Activated Ionisation - MAI); decrease in atomic ions (Molecular Activated Recombination - MAR); or increase in neutral atoms (Molecular Activated Dissociation - MAD) (Tab. \ref{tab:MolCombinations}).

\subsubsection{Diagnosing Detachment}

\begin{figure}[]
    \centering
    \includegraphics[width=0.8\textwidth]{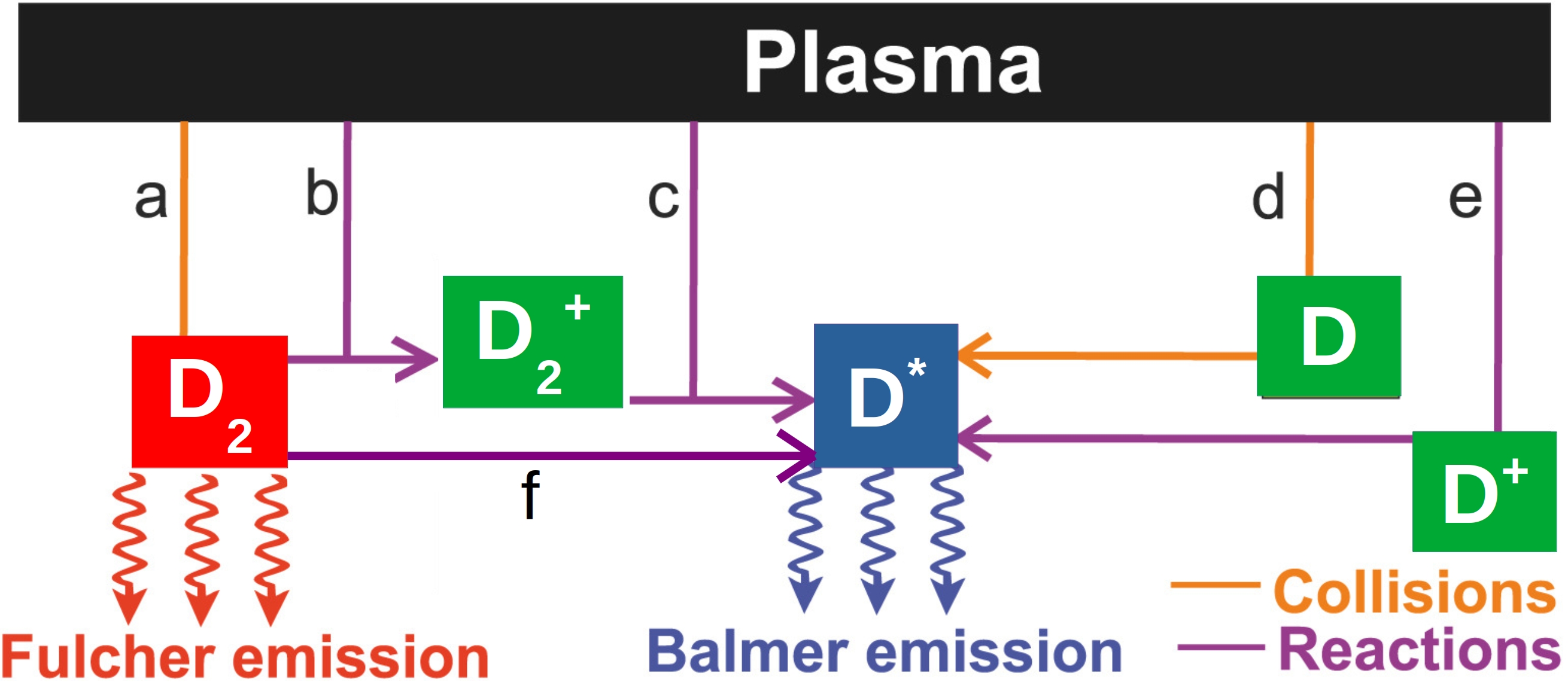}
    \caption{Visual overview of important plasma-molecule and plasma-atom collisions (orange) and reactions (magenta) considered in this work. 
    (a) Collisions between the plasma and $D_2$ excite the molecule electronically resulting in Fulcher line emission; (b) Reactions between the plasma and $D_2$ result in the formation of molecular ions ($D_2^+$); (c) Reactions (MAR, MAD) between the plasma and $D_2^+$, which can result in excited neutral atoms ($D^*$) and Balmer line emission; (d) EIE between electrons and D excite D, resulting in Balmer line emission; (e) EIR reactions with D resulting in excited D and Balmer line emission; (f) Dissociation events may produce excited neutral atoms. Adapted from \cite{verhaegh2021molecule}.}
    \label{fig:EmissionInteractionsUpdate}
\end{figure}

Detailed analysis of the emission from excited hydrogen atoms facilitates studying the microscopic origins of detachment, for instance by analysing the visible Balmer emission spectrum.
Excited atoms, directly produced from EIR, MAR and dissociation events, as well as from electron-impact-excitation (EIE) (a known predecessor for atomic ionisation) can de-excite to the second energy level, releasing emissions of the Balmer series.
Additionally, Fulcher emissions produced from the electronic excitation of deuterium molecules is also visible.
Fulcher emissions give information for detachment via inference of temperature \cite{2022Spectroscopic}.
A summary of visible detachment-related interactions is given in Fig. \ref{fig:EmissionInteractionsUpdate}.

Utilising the intensity of these emissions (denoted by $B$ throughout this report) forms the basis for detachment diagnostics.
Conventional and ultrafast spectrometers within MAST-U deploy line-of-sight (LoS) spectroscopy through the divertor chamber, resulting in a line integrated intensity (in $\mathrm{ph/m^2/s}$) for the individual Balmer lines with an uncertainty of 12.5\%.
The conventional spectrometer, spread over 40 lines-of-sight, can detect $B_\alpha$, $B_\gamma$ and $B_\delta$, whilst the ultrafast spectrometer, spread over 10 lines-of-sight, finds $B_\alpha$ and $B_\beta$.

\begin{table}[]
\centering
\begin{tabular}{c|cccc}
Transition      & $3 \rightarrow 2$ & $4 \rightarrow 2$ & $5 \rightarrow 2$ & $6 \rightarrow 2$ \\
Name            & $B_\alpha$        & $B_\beta$         & $B_\gamma$        & $B_\delta$        \\
Wavelength / nm & 656.28            & 486.14            & 434.05            & 410.17           
\end{tabular}
\caption{Visible spectrum of the Balmer line emission series.}
\label{tab:BalmerEmissions}
\end{table}

There is great difficulty associated with the diagnosis of detachment using Balmer emissions.
All Balmer emissions come from excited atoms, however these same atoms can originate from both atomic and molecular plasma-interactions, with no experimental method to distinguish between contributions from individual interactions.
Additionally, there are no emissive precursors for MAI so its inference relies on the knowledge of the ratio of molecular reactions given in Tab. \ref{tab:MolCombinations}.
Our analysis of detachment is therefore tasked with deducing the many potential interactive origins for the detected Balmer emissions, using only the intensity of these emissions.

\subsubsection{Detachment in the Super-X}

\begin{figure}[]
    \centering
    \includegraphics[width=0.8\textwidth]{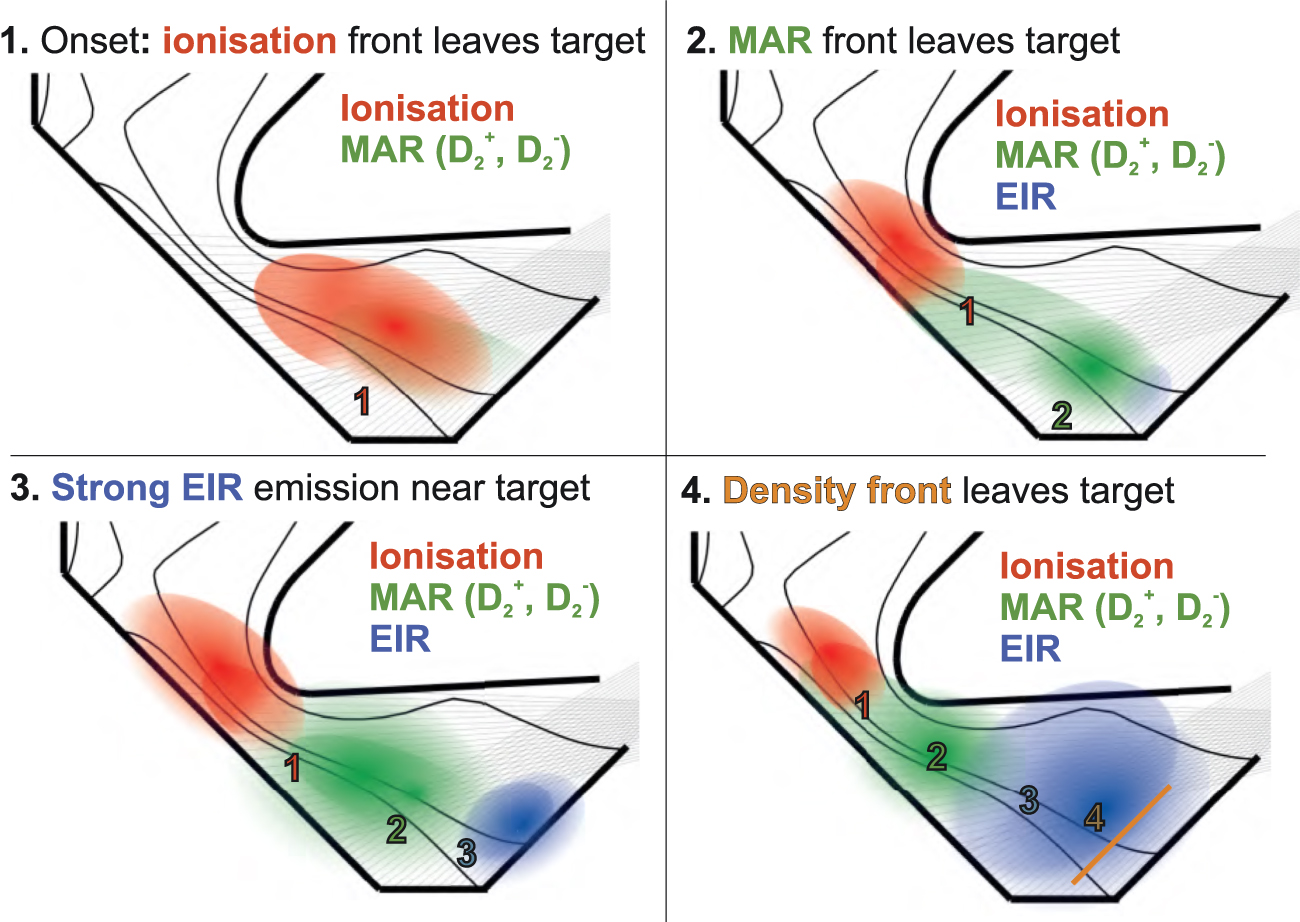}
    \caption{Adopted from \cite{2022Spectroscopic}:
    Schematic overview of the four inferred MAST-U Super-X detachment phases in terms of the reactions occurring in the divertor.
    Also shown is the Super-X plasma geometry and the DMS spectroscopic viewing chords. 
    The numbers shown indicate: (1) the back-end of the ionisation region; (2) the back-end of the molecular activated recombination (MAR) region; (3) the front-end of the electron ion recombination (EIR) region; and (4) the back-end of the electron ion recombination/density region. 
    The magnetic geometry in this illustration has been obtained from a SOLPS-ITER simulation (from \cite{myatra2021numerical}).
    }
    \label{fig:Detachment}
\end{figure}

Experimentation and pre-existing analyses have found that detachment within the Super-X can be split into four distinguishable phases (Fig. \ref{fig:Detachment}) \cite{2022Spectroscopic, wijkamp2023characterisation}:
\begin{enumerate}
    \item The ionisation (both atomic and molecular-activated) region detaches from the target.
    Subsequently, increased molecular densities occur in the space left by ionisation near the target, resulting in increases in MAD and more importantly MAR.
    Significant Balmer line emissions are associated with these increases in plasma-molecule interactions.
    \item Plasma-molecule interactions shift upstream, indicated by the movement of the notable Balmer emissions.
    The increased ion sinks begin to decrease ion, and hence heat, flux downstream.
    \item Temperatures drop below $0.5$eV near the target, providing necessary conditions for EIR to become significant.
    \item Balmer emissions associated with EIR detach from the target, suggesting the movement of the bulk of the electron density upstream.
\end{enumerate}

\subsection{This Work}


\begin{figure}[]
    \centering
    \includegraphics[width=0.8\textwidth]{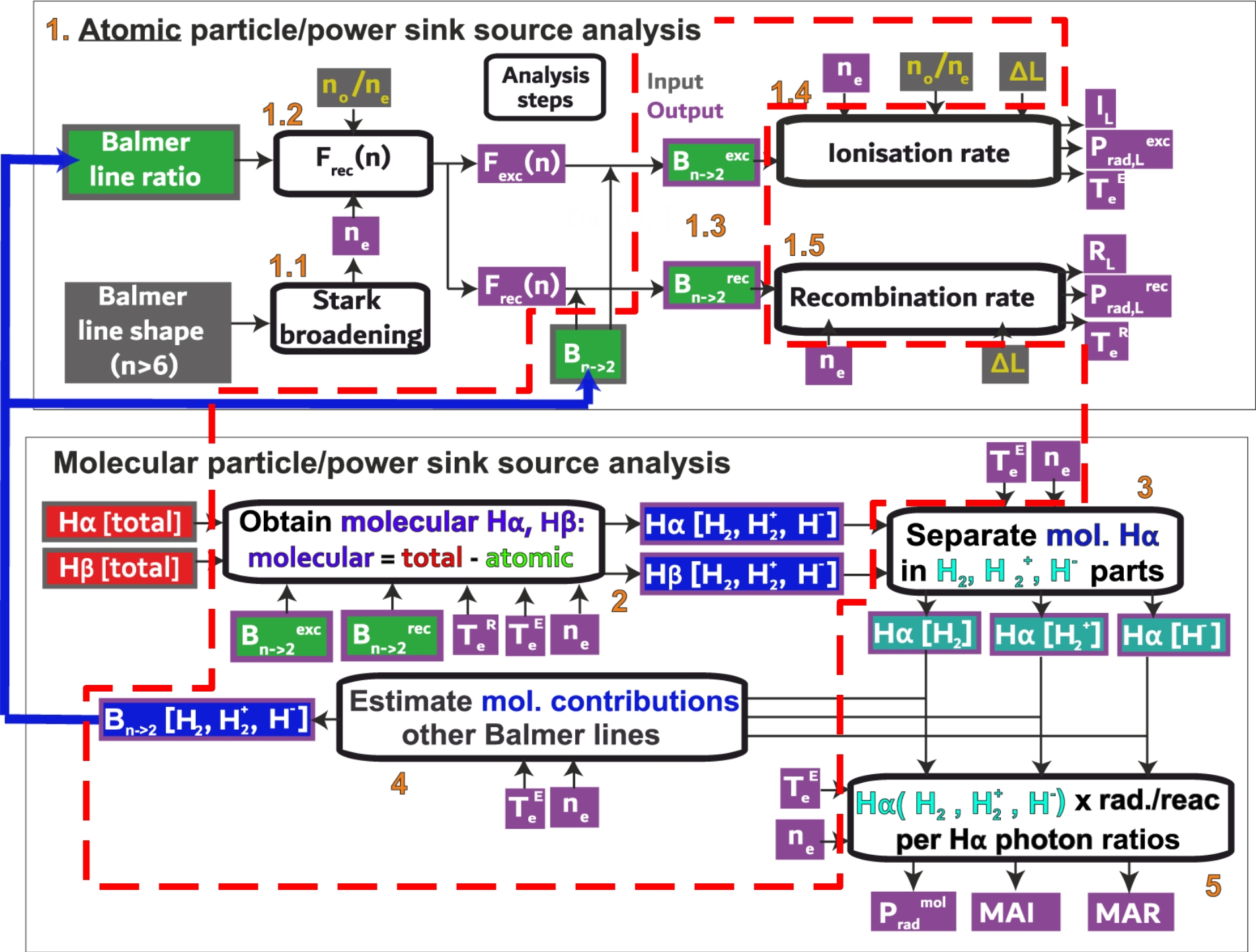}
    \caption{Schematic overview of the full BaSPMI routine. 
    The orange numbers indicate the sequence of the various steps.
    Model parameters: $\frac{n_0}{n_e}$, $\Delta L$, $n_e$, $T_e^E$, $T_e^R$, discussed in Section. \ref{sec:EmissionModel}.
    Dashed-red box: analysis section replaced in BaySPMI and our analysis technique.
    Adopted from \cite{verhaegh2021novel}.}
    \label{fig:BaSPMIUpdate}
\end{figure}

The inference of the detachment stages in \cite{2022Spectroscopic} was performed using BaSPMI \cite{verhaegh2021novel}, a complicated and lengthy iterative process, altering model parameters until a combination of emissive processes and thus plasma interactions have converged.
The BaSPMI algorithm is summarised in Fig. \ref{fig:BaSPMIUpdate}, with note that it is displayed to illustrate the complexity of such an analysis.

Steps have been taken to disentangle this process by utilising Bayesian inference techniques combined with a simplified emission model to replace the section enclosed in the dashed red line in Fig. \ref{fig:BaSPMIUpdate}.
This update, referred to as BaySPMI, uses the emission model to infer the most likely quantities for the free parameters, which can then be used to calculate rates of ion sources and sinks.

Both BaSPMI and BaySPMI require extensive computational resources ($\lesssim$120GB of memory) and time ($\lesssim$48 hours) to complete, making the analysis of detachment largely inaccessible.
With detachment expected to play an essential role for target heat deposition reduction, it is crucial to have a practical analysis method that can be used to evaluate a wide number of detachment scenarios to better understand the physics involved.

Importantly, BaSPMI must have three Balmer emission lines to work.
Therefore, the observations from two Balmer lines diagnosed by the ultrafast spectrometer is insufficient for use of BaSPMI.
The analysis of ultrafast data is essential for investigation into specific sections of shots to study the effect of transient events, such as ELMs, on detachment and divertor physics.

We therefore build on and improve the analysis methods used thus far, specifically BaySPMI, to create a more practical and efficient detachment analysis software.
Success for such a software will be measured by; 
\begin{itemize}
    \item Practicality: ability to run between MAST-U pulses (20 minutes) on a scientific workstation.
    \item Validity: successful reproduction of ideal test case and simulated data and agreement with BaySPMI results.
    \item Ability: able to analyse conventional and ultrafast observations.
\end{itemize}

Once validated, the software is used to analyse a simulated $N_2$ gas puffing experiment, in which $N_2$ is seeded into the divertor near its entrance at rates varying from $1\times10^{21}$ to $1.1\times10^{22}$ $\mathrm{mol./s}$ \cite{myatra2023predictive}.
Discharge 46860 (Tab. \ref{tab:46860}), a density ramp experiment from the second MAST-U campaign, is also analysed.

\begin{table}[h]
\centering
\begin{tabular}{llllll}
Discharge & $I_p$ / kA & $P_\mathrm{SOL}$ / MW & $f_\mathrm{GW}$ / \% & Fuelling  & Description                                                  \\ \hline
46860     & 750        & 1.1 - 1.3      & 25 - 50       & LFS-V & \begin{tabular}[c]{@{}l@{}}Beam heated\\ L-mode\end{tabular}
\end{tabular}
\caption{Table showing main discharge parameters, including plasma current ($I_p$),
power crossing the separatrix into the SOL ($P_{SOL}$), core Greenwald fraction ($f_{GW}$), fuelling location and a description.}
\label{tab:46860}
\end{table}

\section{Methodology}


\subsection{Balmer Emission Model} \label{sec:EmissionModel}

The emission model used in BaySPMI forms the foundation of our analysis.
Splitting the Balmer emission intensity for each emission band into plasma-atom and plasma-molecule interactions structures the diagnostic model:
\begin{equation}
    \begin{split}
        \label{eq:AtomMol}
        B_\mathrm{n\rightarrow2}&=B_\mathrm{n \rightarrow 2}^\mathrm{atom} + B_\mathrm{n \rightarrow 2}^\mathrm{mol}\\
    \end{split}
\end{equation}

Current understanding derives Eq. \ref{eq:AtomMolH2} from Eq. \ref{eq:AtomMol} \cite{2022Spectroscopic}, recognising the individual contributions to Balmer emissions and identifying the six model parameters required to ultimately reconstruct the ion sources and sinks.

\begin{equation}
    \begin{split}
        \label{eq:AtomMolH2}
        B_\mathrm{n\rightarrow2}=&B_\mathrm{n \rightarrow 2}^\mathrm{EIE} + B_\mathrm{n \rightarrow 2}^\mathrm{EIR} + B_\mathrm{n \rightarrow 2}^\mathrm{PMI}\\
        =&\Delta L n_e^2 \frac{n_H}{n_e} \mathrm{PEC}^\mathrm{EIE}_\mathrm{n \rightarrow 2}(n_e, T_e^E) + \Delta L n_e^2 \mathrm{PEC}^\mathrm{EIR}_\mathrm{n \rightarrow 2}(n_e, T_e^R)\\
        &+ \Delta L n_e n_{H_2^+} \mathrm{PEC}^\mathrm{PMI}_\mathrm{n \rightarrow 2}(n_e, T_e^E, T_e^R)
    \end{split}
\end{equation}.

There is a requirement for databases of photon emissivity coefficients (PEC) to compute the model.
Such coefficients identify the amount of photons we expect to see for the respective interactions at a given electron temperature and density.
$\mathrm{PEC}^\mathrm{EIE}_\mathrm{n \rightarrow 2}$ and $\mathrm{PEC}^\mathrm{EIR}_\mathrm{n \rightarrow 2}$ were obtained using ADAS \cite{o2013adas} and $\mathrm{PEC}^\mathrm{PMI}_\mathrm{n \rightarrow 2}$ using Yacora \cite{wunderlich2020yacora}.
All PEC databases were created for temperatures ranging from 0.02 - 50eV and electron densities of $1\times10^{18}$ - $5\times10^{20}\mathrm{m^{-3}}$. 

Before stating the free parameters involved in Eq. \ref{eq:AtomMolH2}, it should first be noted that the density of molecular ions ($n_{H_2^+}$) presents a large unknown that would cause unacceptable uncertainties if included.
Therefore, we use the ratio for plasma-molecular to plasma-atomic emissions for the Balmer-$\alpha$ series (known as $Q_\mathrm{mol}$) as a scaling factor to bypass the use of such an uncertain parameter.
Accommodating for $Q_\mathrm{mol}$ is shown below.

\begin{equation}
    \begin{split}
        B_\mathrm{n \rightarrow 2} =& B_\mathrm{n \rightarrow 2}^\mathrm{atom} + B_\mathrm{n \rightarrow 2}^\mathrm{mol} \\
        =& B_\mathrm{n \rightarrow 2}^\mathrm{atom} + B_\mathrm{n \rightarrow 2}^\mathrm{mol}\left[\frac{B_\mathrm{3 \rightarrow 2}^\mathrm{atom}}{B_\mathrm{3 \rightarrow 2}^\mathrm{atom}}\right]\left[\frac{B_\mathrm{3 \rightarrow 2}^\mathrm{mol}}{B_\mathrm{3 \rightarrow 2}^\mathrm{mol}}\right]\\
        =& B_\mathrm{n \rightarrow 2}^\mathrm{atom} + B_\mathrm{3 \rightarrow 2}^\mathrm{atom}\left[\frac{B_\mathrm{3 \rightarrow 2}^\mathrm{mol}}{B_\mathrm{3 \rightarrow 2}^\mathrm{atom}}\right]\left[\frac{B_\mathrm{n\rightarrow2}^\mathrm{mol}}{B_\mathrm{3\rightarrow2}^\mathrm{mol}}\right]\\
        =& B_\mathrm{n\rightarrow2}^\mathrm{atom} + B_\mathrm{3\rightarrow2}^\mathrm{atom} Q_\mathrm{mol} \frac{B_\mathrm{n\rightarrow2}^\mathrm{mol}}{B_\mathrm{3\rightarrow2}^\mathrm{mol}}\\
        =& B_\mathrm{n \rightarrow 2}^\mathrm{EIE} + B_\mathrm{n \rightarrow 2}^\mathrm{EIR} + Q_\mathrm{mol}\left[B_\mathrm{3 \rightarrow 2}^\mathrm{EIE} + B_\mathrm{3 \rightarrow 2}^\mathrm{EIR}\right]\left[\frac{B_\mathrm{n \rightarrow 2}^\mathrm{PMI}}{B_\mathrm{3 \rightarrow 2}^\mathrm{PMI}}\right]\\
    \end{split}
\end{equation}

Taking a ratio of the PMI's allows the cancellation of $n_{H_2^+}$ making our final model for Balmer line emissions less sensitive to this uncertainty.

\begin{equation}
    \begin{split}
        \label{eq:EmissionModel}
        B_\mathrm{n \rightarrow 2} = \Delta L n_e^2 &\Biggr[\frac{n_H}{n_e} \mathrm{PEC}^\mathrm{EIE}_{n \rightarrow 2}(n_e, T_e^E) + \mathrm{PEC}^\mathrm{EIR}_\mathrm{n \rightarrow 2}(n_e, T_e^R)\\
        &+ Q_\mathrm{mol}\left[\frac{n_H}{n_e} \mathrm{PEC}^\mathrm{EIE}_\mathrm{3 \rightarrow 2}(n_e, T_e^E) + \mathrm{PEC}^\mathrm{EIR}_\mathrm{3 \rightarrow 2}(n_e, T_e^R)\right]\\
        &\left[\frac{\mathrm{PEC}^\mathrm{PMI}_\mathrm{n \rightarrow 2}(n_e, T_e^E, T_e^R)}{\mathrm{PEC}^\mathrm{PMI}_\mathrm{3 \rightarrow 2}(n_e, T_e^E, T_e^R)}\right] \Biggr]
    \end{split}
\end{equation}

Our free parameters required to diagnose the sources and sinks of ions are as follows: path length ($\Delta L$), density of electrons ($n_e$), ratio of neutral hydrogen ($\frac{n_H}{n_e}$, referred to as $n_\mathrm{ratio}$ throughout the rest of this report), the temperature for the population of electrons along the LoS that could experience EIE and EIR ($T_e^E$ and $T_e^R$ respectively) and $Q_\mathrm{mol}$.

The path length parameter identifies the extent of plasma that the LoS intersects and is therefore the length in which the spectrometer integrates to produce the measured line-integrated intensity.
$n_\mathrm{ratio}$ is used rather than $n_H$, as it introduces the understanding of the relationship between $n_e$ and $n_H$ into the model.
During our analysis $Q_\mathrm{mol}$ is interchanged with a similar parameter: $f_\mathrm{mol}$.
The relation between the two parameters is given in Eq. \ref{eq:fmol}.
$f_\mathrm{mol}$ is more practical during analysis as it can only have values between 0 and 1.

\begin{equation}
\label{eq:fmol}
\begin{aligned}
    f_\mathrm{mol} &= \frac{B_\mathrm{3 \rightarrow 2}^\mathrm{mol}}{B_\mathrm{3 \rightarrow 2}^\mathrm{mol} + B_\mathrm{3 \rightarrow 2}^\mathrm{atom}}, &\quad
    Q_\mathrm{mol} &= \frac{B_\mathrm{3 \rightarrow 2}^\mathrm{mol}}{B_\mathrm{3 \rightarrow 2}^\mathrm{atom}}, &\quad
    Q_\mathrm{mol} &= \frac{f_\mathrm{mol}}{1 - f_\mathrm{mol}}
\end{aligned}
\end{equation}

By utilising an emission model that only requires the Balmer-$\alpha$ and one additional line emission intensity, we are able to analyse measurements from the ultrafast spectrometer, providing an important precursor for the success of our analysis.

Ultimately, the analysis of interest is how the various plasma-atom/molecule interactions impact power and particle balance in the divertor. 
Using the post-processing developed in \cite{verhaegh2021novel}, the various emission fractions of the Balmer line emission (EIE, EIR, PMI) can be used to estimate the ion sources/sinks and power losses from each of these processes using the six free parameters introduced in the Balmer line emission model.

\subsection{Bayesian Analysis}

\begin{figure}
    \centering
    \includegraphics[width=0.8\textwidth]{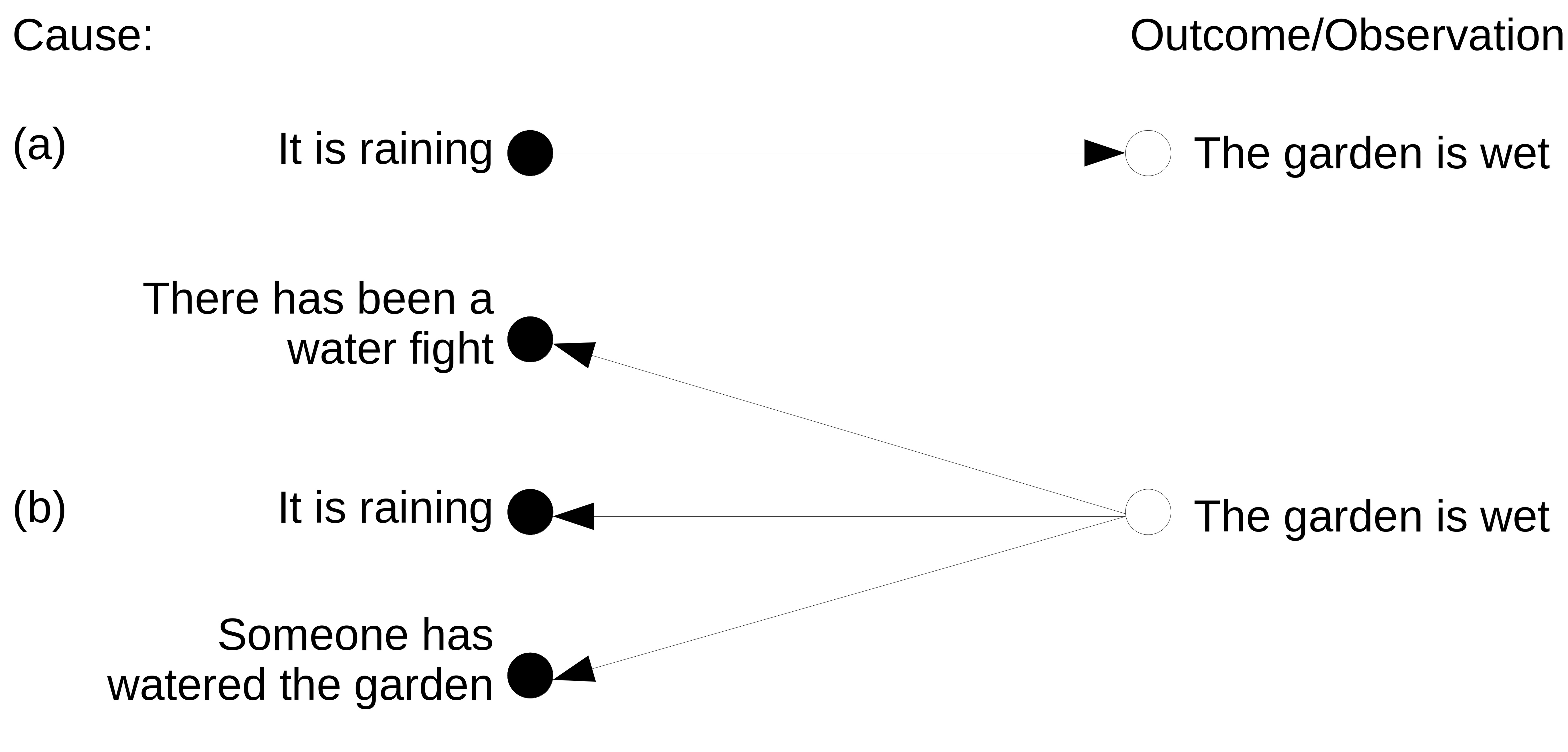}
    \caption{Schematic example of deductive logic (a) vs plausible reasoning (b). Adapted from \cite{2006Data}}
    \label{fig:DeductiveVsReason}
\end{figure}

Much of the difficulties of experimental data analyses can be reduced to the inherent reversal of deductive logic.
Deductive logic follows that: given a cause, an outcome can be deduced, the reverse of this being inductive logic: given an outcome, what is the original cause?
The complexities of inducing causes are indicated in Fig. \ref{fig:DeductiveVsReason}, with there being many possible causes for any one given observation.
Indeed, this is the case for our analysis of detachment in Eq. \ref{eq:EmissionModel}: given the observation of the intensity of Balmer emissions, what combination of parameter values would have caused this?

Often, the value of the parameters within the best matching model are taken to be fixed with some level of uncertainty, this is the approach used in BaSPMI.
Moving away from convention, Bayesian probability theory provides a powerful alternative approach towards data analysis, replacing fixed parameter natures with a probability distribution of all possible parameter values.
In doing so, the distribution of all possible causes can be inferred \cite{bowman2020development}, where fixing parameters in other analysis techniques may fail to identify plausible causes.

\subsubsection{Bayes' Theorem}

The notation used in probability theory is as follows. 
The probability (P) that X is true: P(X).
The probability that both X and Y are true can be denoted using a comma: P(X, Y).
Conditional probabilities are represented using `|', for example, the probability that X is true given that Y is true: P(X|Y).

Bayes' theory provides logical reasoning for data analysis by stating the following formula:
\begin{equation}
\label{eq:Bayes}
    P(X|Y) = \frac{P(Y|X) \times P(X)}{P(Y)}
\end{equation}

Whilst fairly unassuming in theory, applying it in the context of science showcases its significance for data analysis \cite{2006Data}:
\begin{equation}
        P(\mathrm{hypothesis|data}) = \frac{P(\mathrm{data|hypothesis}) \times P(\mathrm{hypothesis})}{P(\mathrm{data})}   
\end{equation}
\begin{equation}
\label{eq:ScienceBayes}
    P(\mathrm{hypothesis|data}) \propto P(\mathrm{data|hypothesis}) \times P(\mathrm{hypothesis})
\end{equation}

This project omits the need of P(data), formally named `evidence', throughout.
By representing the probability that the data is true within a single experiment makes such a term a normalised constant which can therefore be excluded whilst still producing comparable results \cite{bowman2020development}.

For the remainder of this report the other components of Bayes' theorem will also be referred to with their formal names.
From right to left in Eq. \ref{eq:ScienceBayes}; the prior probability; likelihood function; and posterior probability.
The prior accumulates previous knowledge of the experiment to form a probability relating to the current believed truth of the hypothesis.
The likelihood function utilises an aforementioned model to compose a probability that the data is correct in the context of the hypothesis.
Lastly, the posterior combines the prior and likelihood to produce the probability that the hypothesis is true in light of the experimental data.

Thus, the nature of Bayesian statistical analysis is to methodically test many hypotheses to generate an evolving probability density function (PDF) of the posterior outcome for said hypotheses, with areas of high probability density signifying likely hypotheses.

\subsubsection{Bayesian Analysis for Detachment}

Given our experimental data of emission intensities for Balmer lines ($B^\mathrm{exp}_\mathrm{n\rightarrow2}$), the hypotheses proposed are the combinations of free parameters that contribute to the emission model in Eq. \ref{eq:EmissionModel} ($T_e^E, T_e^R, n_e, n_\mathrm{ratio}, f_\mathrm{mol}, \Delta L$).
Therefore, the application for Bayes' theorem in regards for this project is the following:
\begin{equation}
\begin{split}
    P(T_e^E, T_e^R, n_e, n_\mathrm{ratio}, f_\mathrm{mol}, \Delta L|B^\mathrm{exp})
    =  &\prod^n P(B^\mathrm{exp}_\mathrm{n\rightarrow2}|T_e^E, T_e^R, n_e, n_\mathrm{ratio}, f_\mathrm{mol}, \Delta L \rightarrow B^\mathrm{pre}_\mathrm{n\rightarrow2})\\
    & \times P(T_e^E, T_e^R, n_e, n_\mathrm{ratio}, f_\mathrm{mol}, \Delta L)
\end{split}
\end{equation}
        
\textit{Prior Probability:}
Our priors, ($P(T_e^E, T_e^R, n_e, n_\mathrm{ratio}, f_\mathrm{mol}, \Delta L)$), assemble knowledge of the values for these parameters.
For example, if a parameter hypothesis is proposed with $T_e^E < T_e^R$, then the prior would return an exceptionally low probability, reflecting that we would not expect the $T_e^R$ to ever be greater than $T_e^E$, since EIE occurs at higher temperatures than EIR.
Since the same values for our free parameters are used within one evaluation of numerous Balmer emission bands, they need only be applied once.

\textit{Likelihood Function:} 
Assuming such hypothesised parameter values are true, computing the emission model then returns `predicted' emission intensities for each Balmer line, $P(B^\mathrm{exp}_\mathrm{n\rightarrow2}|T_e^E, T_e^R, n_e, n_\mathrm{ratio}, f_\mathrm{mol}, \Delta L \rightarrow B^\mathrm{pre}_\mathrm{n\rightarrow2})$.
Lower likelihood functions are subsequently computed when the experimental results are further in value to the predicted emissions.
The product of the likelihood probabilities across all emission bands produce a final likelihood.
Greater confidence in the resultant posterior is associated with increased Balmer lines introduced in the analysis, since more individual likelihood functions must agree on any given combination of parameters, with the exception for if our emission model was imperfect.

\textit{Posterior Probability:}
Knowing that the experimental values are always true, low likelihoods infer low posterior probabilities ($P(T_e^E, T_e^R, n_e, n_\mathrm{ratio}, f_\mathrm{mol}, \Delta L|B^\mathrm{exp})$) for the current combination of parameters due to their proportionality in Bayes' theorem.
Therefore, our posterior forms a function which tells us how likely it is that a given set of parameter values were the cause of the observed data.

A successful way of systematically investigating varying compositions of free parameters is to set up a grid, of dimensions equal to that of the number of variables, with each `cell' on this grid composed of specific combinations of values for these parameters.
This `cell' is then evaluated to compute the posterior probability for that parameter combination.
The probability of each cell directly contributes to the total PDF for the combined parameter space.
BaySPMI creates such a grid, of 144,000,000 cells spread over the space of realistic values, to perform its analysis of detachment.
Whilst the parameter values in their individuality are all plausible, within six parameter space, the majority of their combinations are not.
Thus many evaluations are wasted on `cells' that provide no realistic probability, making the process very inefficient.
Additionally, all 144,000,000 evaluations are stored in memory during application causing large computational strain.
Overcoming this inefficiency is where our analysis differs from BaySPMI.

For information on the six output parameters to be used in the post-processing calculations to infer ion sources/sinks, the calculations need to be computed on a set of sample parameters, sampled from the multidimensional PDF to account for all the various correlations.
Therefore, sampling is performed with bias based upon where the combination of parameters for each cell lies probability-wise within the PDF.
Parameter combinations with higher probability density are more likely to be sampled, whilst combinations with lower probability densities are less likely.
The representation of probability within the sampling is a powerful method of uncertainty propagation within Bayesian analyses.
Because these weighted samples are carried forward in processing, probabilities for the original parameters continue to hold weight throughout computation.
Thus, a sample of new computed parameters is created (for example, a sample for the ionisation source), which can be mapped back to a PDF using kernel density estimation.

\subsection{Adaptive Grid}

\begin{figure}
    \centering
    \includegraphics[width=0.85\textwidth]{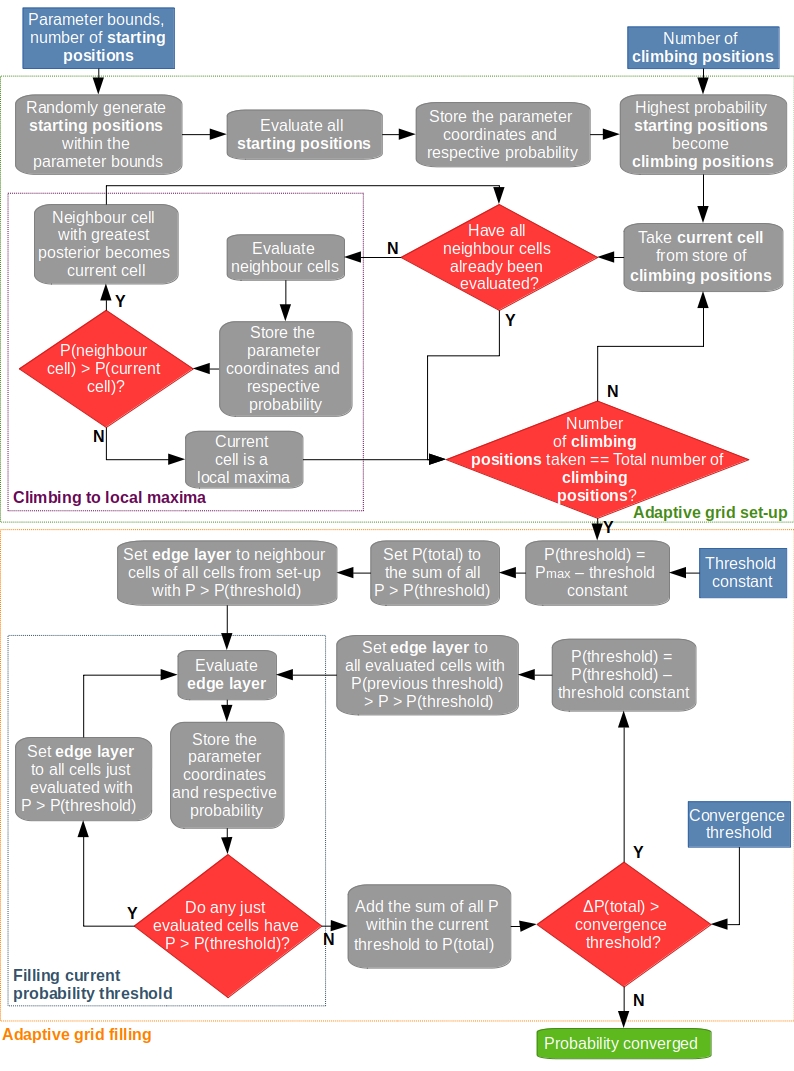}
    \caption{Blue: Inputs into algorithm. Grey: Computational process. Red: Decisions made by the algorithm. Green: Final output. Dashed lines represent the different phases of the algorithm. Contents adopted from \cite{2016Bayesian}. Infographic adapted from \cite{2019Bayesim}.}
    \label{fig:AdaptiveFlowchart}
\end{figure}

The grid technique in BaySPMI can be referred to as `full grid', since it evaluates the full parameter space.
We apply a new algorithm that evaluates only the areas of parameter space with high probability, preventing the computation of unrealistic parameter combinations.
Such a technique is referred to as an `adaptive grid' approach.
A conceptual overview of the adaptive grid algorithm, initially created in \cite{2016Bayesian} and updated for use in this project, is given in Fig. \ref{fig:AdaptiveFlowchart}.

Adaptive grid analyses pairs very well with Bayesian inference techniques, as the computed probability can inform the script whether an area in parameter space is worth exploring.
Additionally, working in probability space allows confidence that all useful parameter space has been explored by examining the convergence in total probability found thus far.
For example, the convergence threshold in Fig. \ref{fig:AdaptiveFlowchart} is set to be sufficiently low, that when continuations of the adaptive grid algorithm fail to add probability greater than this small value, we can be sure all important parameter space has been explored.
To ensure we do not converge in probability by exploring only one local maxima, we expand our number of starting positions to be sufficiently large such that we find and explore the majority of maxima.

In addition to the algorithm inputs shown in Fig. \ref{fig:AdaptiveFlowchart}, the adaptive grid script requires the definition for the desired intervals between parameter cells and a centre of the grid.
This centre is not used as a starting position, but as a method of defining an offset such that each cell can be given real parameter values.

The decision to use an adaptive grid approach for more efficient analysis of parameter space over more established statistical techniques such as Markov Chain Monte Carlo (MCMC) \cite{andrieu2008MCMC} was made due to the random nature of MCMC.
Analyses with reduced amounts of randomness, such as our adaptive grid technique, produce more deterministic results.

Successful application of an adaptive grid technique, differing in approach to ours, has been demonstrated for up to four dimensions in parameter space \cite{2019Bayesim}.
Whilst shown to be significantly faster than full grid approaches, notably coarse cell spacing and reduced exploration of parameter space was required to make such an analysis feasible in four dimensions without the requirement of a high performance computer (HPC).
We will use our alternative technique to evaluate an increased number of dimensions (six), with finer cell spacing and as much accuracy as its full grid counterpart, in significantly less time, without the requirement of a HPC.

\section{Software Development}

\subsection{Application for Bayesian Analysis in Script}

\begin{figure}
    \centering
    \includegraphics[width=0.7\textwidth]{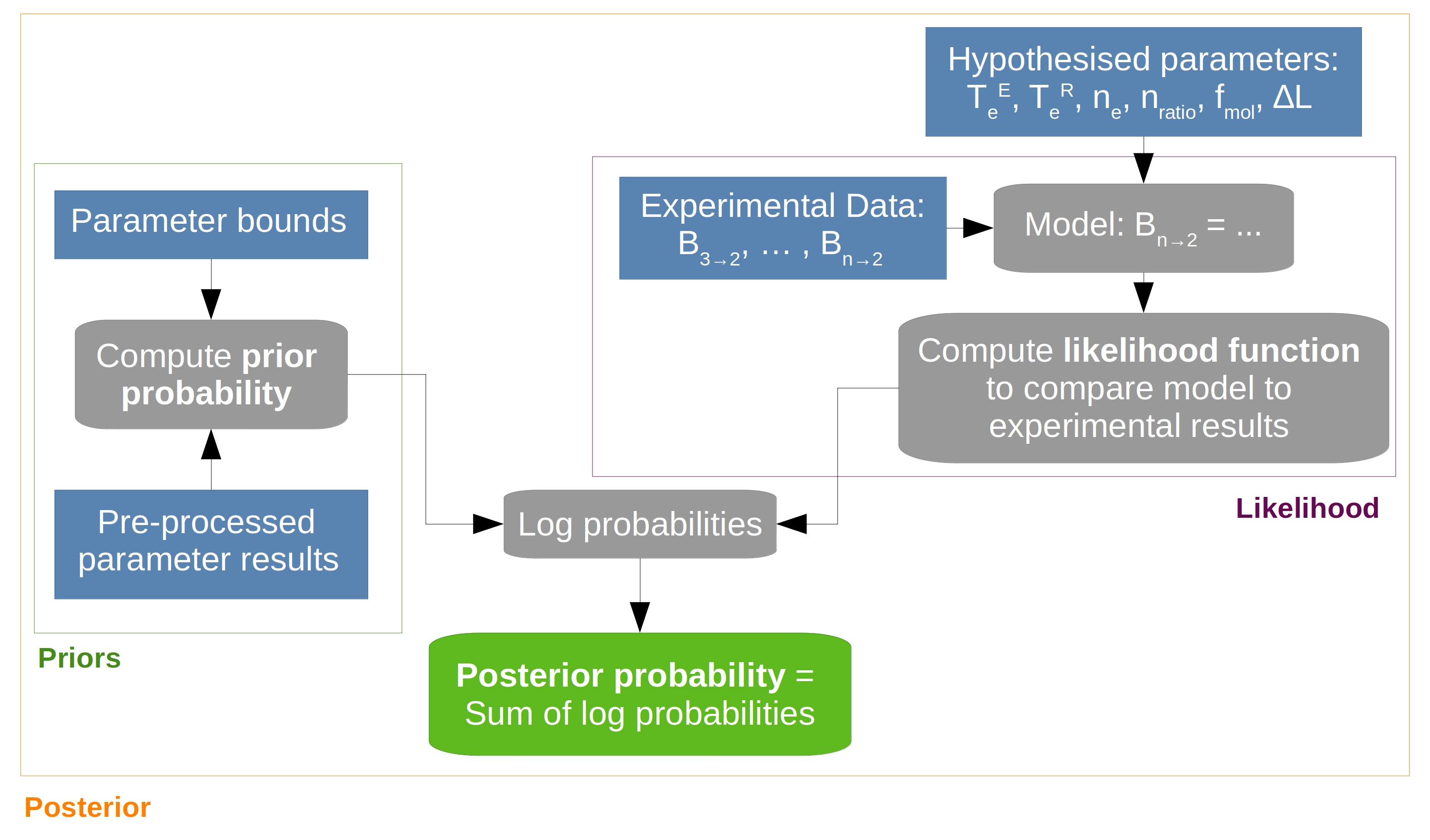}
    \caption{Blue: Inputs into evaluation. Grey: Computational process. Green: Final output. Continuous lines separating the different components of Bayes' theory represent the modular set up within the script.}
    \label{fig:BayesApplication}
\end{figure}

Given in Fig. \ref{fig:BayesApplication} is the relation between components of Bayes' theorem within our analysis.
Our hypothesised parameters are the values for our free parameters within the current cell.
The prior module is discussed in the following section.
We use the program MIDAS (used in \cite{bowman2020development}) to compute the Balmer line emissions within our model.
Our incorporation of Bayes' utilises the log-probability environment.
Working in log-probability space: simplifies computation via addition of probabilities rather than multiplying, ensures probabilities are always positive and enforces unrealistic probability space via the addition of large negative numbers.

Within our parameter space created for the Bayesian analysis, we are using the log values for $T_e^E$, $T_e^R$, $n_e$ and $n_\mathrm{ratio}$, then converting these back to their actual values during sampling.
This is done because it allows equally weighted interval's in parameter spaces spanning across multiple magnitudes.

\subsubsection{Implication of Priors}\label{sec:PriorImplication}

The utilisation of prior probabilities is what allows an analysis of so many parameters from such little input data.
Without these priors, the parameter space would be left unrestricted, making the completion of the adaptive grid algorithm and the production of valid results largely impossible.
Our prior implementation can be split into two approaches: hard parameter boundary limits and model parameter value influence.

Hard parameter boundaries define the limits of realistic parameter values seen in our application of the analysis software.
Hence, any hypotheses proposed with a parameter value outside that of the limits shown in Tab. \ref{tab:HardBounds} will have a large, negative log-probability enforced.
This ensures the resultant posterior is highly improbable.
The aforementioned $T_e^E$ < $T_e^R$ prior probability is the other hard limit prior utilised in our analysis.

\begin{table}[h]
\centering
\begin{tabular}{l|cccccc}
            & $T_e^E$ / eV & $T_e^R$ / eV & $n_e$ / $\mathrm{m^{-3}}$ & $n_\mathrm{ratio}$      & $f_\mathrm{mol}$ & $\Delta L$ / m \\ \hline
Lower bound & 0.20         & 0.02         & $5\times10^{18}$ & $1\times10^{-3}$ & 0.01      & 0.02           \\
Upper bound & 5.00         & 1.00         & $4\times10^{19}$ & 1.00             & 0.95      & 0.40          
\end{tabular}
\caption{Parameter limits used for this analysis as well as in BaySPMI.}
\label{tab:HardBounds}
\end{table}

Limits for $T_e^E$, $T_e^R$ and $n_\mathrm{ratio}$ are defined via knowledge of plasma chemistry within the divertor.
Detachment is accessed below $\approx$5eV, EIR below $\approx$1eV and a $n_\mathrm{ratio}$ of 1 states that half the plasma is neutrals, more neutrals than this would fall into unrealistic plasma conditions.
Temperatures below 0.2eV fall outside of EIE conditions, 0.02eV represent the absolute minimum temperature plausible in a divertor and a $n_\mathrm{ratio}$ of $1\times10^{-3}$ is enough to imply a fully ionised plasma.

$n_e$, $\Delta L$ and a more specific $n_\mathrm{ratio}$ limit is dependent on the individual tokamak and divertor system.
$n_e$ and $\Delta L$ are pre-defined with the values in Tab. \ref{tab:HardBounds} as these are known to be realistic for MAST-U.
However, Stark broadening techniques and magnetic mapping of the divertor can provide better estimates for the limits of $n_e$ and $\Delta L$ respectively for individual shots across different tokamaks/divertors.
These processes can be performed prior to the use of our analysis, thus, if defined in the input data dictionary, our script can swap the pre-defined limits for $n_e$ and $\Delta L$, as well as $n_\mathrm{ratio}$, to more appropriate limits.

$f_\mathrm{mol}$ is set between 0.01 and 0.95 to prevent exceptionally small or large values for $Q_\mathrm{mol}$.
All parameter boundaries are the same as those used for the edge of parameter space in BaySPMI.

Where possible, we use models for the observed parameters that are constructed before our analysis.
The parameter models influence the prior probability given to our free parameters.
Model priors are utilised for $T_e^H$, $n_e$ and $\Delta L$ within our script.
The same diagnostics used to estimate the parameter limits for $n_e$ and $\Delta L$ also provide statistical averages of the parameters, used to construct models.
Measured intensity for the Fulcher band emissions across the divertor can be modelled to provide a prior for $T_e^H$.

\subsection{Sampling Method}

BaySPMI utilises rejection sampling, in which a wider and geometrically simpler sample region is established around the complex PDF.
The samples found to be within the PDF are accepted, whilst samples between the wider sample region and PDF are rejected \cite{casella2004rejection}.
Although functional, this technique can prove inefficient.

Therefore, we use an alternative sampling method.
We perform bias sampling on parameter cells rather than sampling parameter values themselves, resulting in far quicker computation.
Reflecting our original PDF within our sample is very important for uncertainty propagation, as the distribution in probability infers uncertainty.
Hence, bias is given by the normalised probability for each of the cells.
Consequently, a single cell may be sampled many times if it is of high probability, producing the same parameter values each time.

Repeated parameter values do not imitate real samples, therefore, we perform nearest neighbour interpolation by adding/subtracting a randomly generated quantity to each parameter value of the sampled cell.
The values of these added/subtracted quantities do not exceed half the defined cell spacing to ensure we stay within the same cell in parameter space.
We can therefore assume the probability of this randomly generated nearest neighbour to be the same as that of the originally sampled cell.
However, if a cell is sampled on the edge of parameter space the inclusion of randomness can cause the sampled value to exit our parameter bounds.
Subsequently, re-sampling is introduced to the script until all sampled values are within our defined limits.

\subsection{Adaptive Grid Optimisation}


Application for the adaptive grid algorithm was tailored for optimal balance of computational resource use and validity of results.

The spaces defined between parameter values is of high importance for such optimisation.
Spacings used in our software were validated against an adaptive grid analysis utilising far finer cells.
Comparison to parameter spaces in BaySPMI is given in Tab. \ref{tab:CellSpacing}.
Our analysis investigates parameter space at a higher resolution than BaySPMI, putting favour towards the adaptive grid technique.
Cell spacings that are too coarse were seen to prevent the correct convergence of the adaptive grid algorithm.

\begin{table}[h]
\centering
\begin{tabular}{l|cccccc}
\multirow{2}{*}{} & \multicolumn{6}{c}{Parameter intervals}             \\ 
                  & $\ln(T_e^E)$ & $\ln(T_e^R)$ & $\ln(n_e)$ & $\ln(n_\mathrm{ratio})$ & $f_\mathrm{mol}$ & $\Delta L$   \\ \hline
BaySPMI            & 0.08    & 0.10    & 0.15   & 0.36      & 0.07 & 0.02 \\
Adaptive Grid     & 0.06    & 0.09    & 0.09   & 0.15      & 0.06 & 0.02
\end{tabular}
\caption{Parameter values between each `cell'.
Units are the same as in Tab. \ref{tab:HardBounds}, but with the natural log of the unit where necessary.}
\label{tab:CellSpacing}
\end{table}

We define the convergence limit for the script to be $1\times10^{-1}$, as it is shown to be sufficient to investigate all useful parameter space.
Running the script past this limit therefore adds no significant importance to our analysis.

The efficiency of our script means that substantially fewer evaluations need to be performed and stored in memory at any one time, resulting in the ability to distribute our evaluations of the divertor at different lines-of-sight and times across different processes, without risk of overloading the memory, further increasing the speed of analysis.

Our script for the analysis software is given in App. A.
\section{Results}

\subsection{Validation of Software}

\subsubsection{Ideal Test Case}

Our initial test of validity was a simple reproduction of manually created observations (creation discussed in App. B.1).
The first ideal test scenario uses model parameters to produce emission intensities for three Balmer lines: $B_\alpha$, $B_\gamma$ and $B_\delta$, simulating data from conventional spectroscopic measurements.
Of the two types of spectroscopic measurements, this is the better restricted scenario for the adaptive algorithm, since we have three Balmer line emission intensity inputs.
We would therefore expect better agreement with the original parameters within this scenario compared to when we have less Balmer lines.

\begin{figure}
    \centering
    \includegraphics[width=0.9\textwidth]{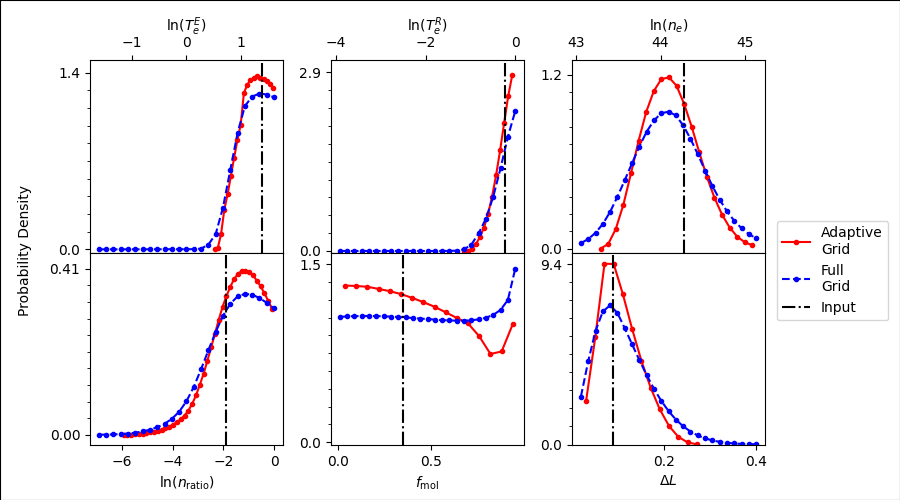}
    \caption{PDFs produced via adaptive (straight-red) and full (dashed-blue) analyses, of Balmer-$\alpha$, $\gamma$ and $\delta$ emissions, for the output parameter values. 
    Probabilities were normalised with respect to their integral (total area associated with the PDF).
    As such, areas under the PDF represent the probability for any given range of parameter values.
    Points on the PDFs display the grid spacings.
    Vertical dashed-dotted lines show the original input parameters.}
    \label{fig:ThreeBandSynParams}
\end{figure}

Accordance of the adaptive grid analysis output with the ideal test case data is seen across all free parameters via an area of high probability corresponding with the input value (Fig. \ref{fig:ThreeBandSynParams}).
Agreement is best for $\Delta L$, due to the incorporated prior model strongly influencing the result of the parameter.
Neglecting prior models and specific boundaries could explain the slightly weaker correspondence between peaks in the PDF and the input value for the density parameters, however the agreement remains good.
A peak in probability is not found for $T_e^R$, this would imply temperatures are too high for recombination to occur, a very likely physical outcome given the relatively high value chosen for $T_e^E$.
Therefore, the inability to produce a peak in probability for $T_e^R$ provides little concern.

Across all parameters mentioned thus far, affinity between adaptive and full grid approaches is very strong.
Such correspondence between the two approaches provides strong confirmation in the ability of the adaptive grid algorithm to analyse the majority of the relevant parameter space.
Differences are seen for $f_\mathrm{mol}$ as the parameter increases in value, however, both approaches are agreeable in probability value for the parameter input.
Overall, the PDF for $f_\mathrm{mol}$ produced by the adaptive algorithm is better biased towards the input value.

\begin{figure}
    \centering
    \includegraphics[width=0.9\textwidth]{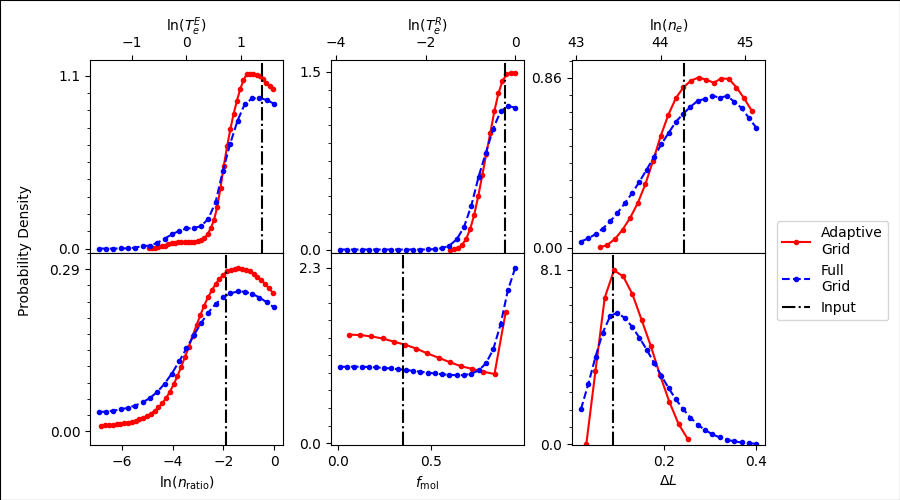}
    \caption{PDFs produced via adaptive (straight-red) and full (dashed-blue) analyses, of Balmer-$\alpha$ and $\beta$ emissions, for the output parameter values.
    Probabilities were normalised via division by integral of PDF.
    Points on the PDFs display the grid spacings.
    Vertical dashed-dotted lines show the original input parameters.}
    \label{fig:TwoBandSynParams}
\end{figure}

With consistent agreement across parameter and emissions (App. B.2) seen in the ideal test case for conventional spectroscopic data, the same analysis is performed with modelled emissions for $B_\alpha$ and $B_\beta$, simulating ultrafast spectroscopic observations.
With only two Balmer lines needing to agree on any given parameter combination the parameter space is less restricted.
We have used the same input parameters as in our previous case, so we expect to see the same overall shapes and locations for our parameter PDFs.
Difficulties seen reproducing these PDFs could infer that less probabilistic restrictions do hinder our adaptive grid analysis.

Fig. \ref{fig:TwoBandSynParams} is significant because it demonstrates the ability for our analysis software to analyse ultrafast spectroscopy data, surpassing the limitations of BaSPMI.
Additionally, much of the same of Fig. \ref{fig:ThreeBandSynParams} is seen in Fig. \ref{fig:TwoBandSynParams}, so a reduced number of input emissions does not greatly impact the accuracy of our analysis at an ideal test case level.
A better peak in $T_e^R$ is now seen.
$f_\mathrm{mol}$ provides a noticeable difference between the two plots, with the adaptive grid providing a similar peak in probability seen in the full grid analysis towards our parameter limits.
Whilst this peak is not correct, the bulk of our probability density remains towards the centre near the input parameter.

Our analysis demonstrates its functionality in reproducing compelling model parameters with minimal inputs, using only knowledge of the hard boundary priors and Balmer emission intensities.

\subsubsection{Simulated Ultrafast Spectroscopic Data}

SOLPS-ITER \cite{2015SOLPS} was used to simulate a scenario within a Super-X-like divertor in which detachment is facilitated by an increasing gas puffing rate of $N_2$, as diagnosed by the ultrafast spectrometer with only $B_\alpha$ and $B_\beta$ measurements.
A constant fuelling rate within the main chamber is chosen such that the density increase caused by fuelling does not facilitate detachment itself \cite{myatra2023predictive}, thus the synthetic data only explores the effects of $N_2$ seeding.
The simulated diagnostic result is obtained through lines-of-sight going through realistic 2D maps of density and temperature.
Therefore, unlike our ideal test case, our simplified model is no longer identical to the analysis input by definition.
However, similar PEC data and assumptions for emission processes are used to create the synthetic data.
With less generalised inputs into our analysis, we utilise all priors discussed in Section \ref{sec:PriorImplication}.

\begin{figure}[]
    \centering
    \includegraphics[width=0.9\textwidth]{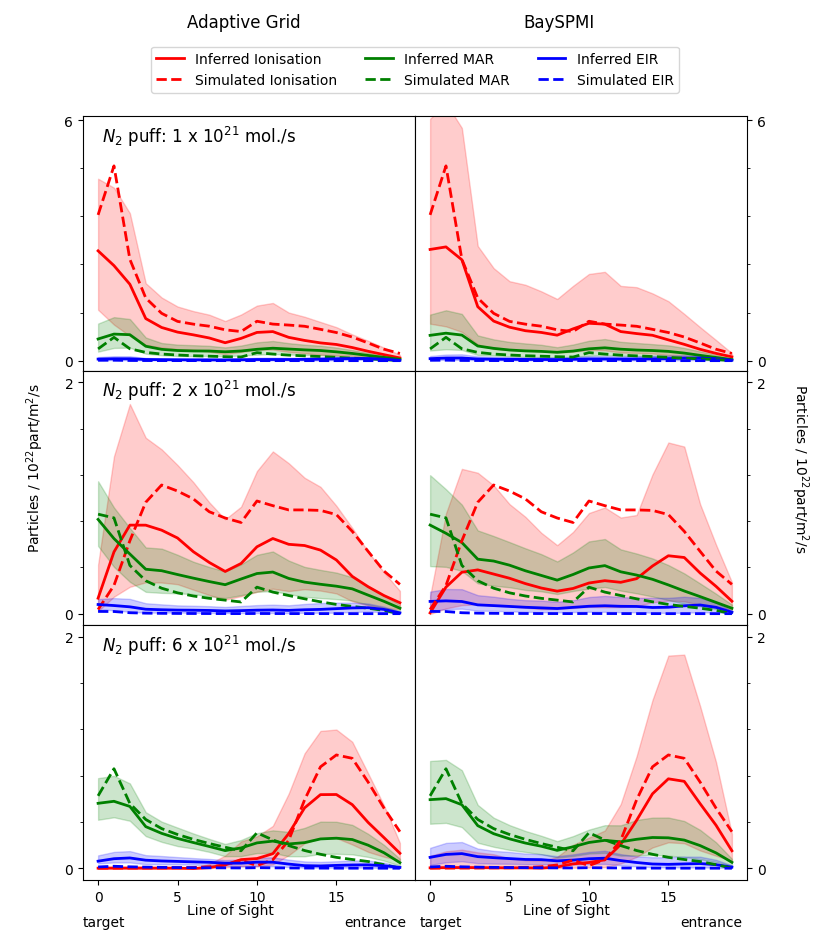}
    \caption{Ion source and sink profiles from analysed results of measurements from an $N_2$ gas puffing simulation in a Super-X-like divertor system, using our adaptive grid analysis (left) and BaySPMI (right).
    Sources and sinks are given as a function of LoS, which can be roughly used as a measurement for distance from target, and measured by number of particles undergoing these interaction at a given area associated with the LoS and `time' point within the simulation.
    Simulated inputs are given in dashed lines whilst inferred outputs are given in full lines.}
    \label{fig:SimSourceSinkCompare}
\end{figure}

We performed the adaptive grid analysis, as well as BaySPMI, to produce free parameter samples, which used the same post-processing techniques to provide inferred measurements for the ion sources and sinks.
Comparison of key free parameters were performed prior to post-processing to verify our analysis was similar to BaySPMI (App. C).
Our adaptive grid analysis is therefore compared against both the simulation inputs and BaySPMI analysis results in Fig. \ref{fig:SimSourceSinkCompare}.

The post-processing takes our sampled combination of free parameters to compute a sample, of equal size, for the ion sources and sinks.
Thus the final observations in Fig. \ref{fig:SimSourceSinkCompare} remain reflective of our original PDF of parameter combinations.
We use the median result of this sample to plot ion sources and sinks, with uncertainties covering the 16-84\% quantiles, representing $\pm$1 standard deviations.
This quantile method has been historically used for BaSPMI and BaySPMI results.

Ionisation is plotted with contribution from both atomic-ionisation and MAI, since we care about the size and location of the ion sources.
MAR and EIR are kept separate due to the information they provide about detachment individually.
MAR is plotted using only the simulated and inferred contributions from $D_2^+$, since it is inferred to be the dominant ion sink \cite{verhaegh2021molecule, 2021Molecular}.
MAD is not included in our results because it does not directly impact the ion sources and sinks related to our power and particle analysis of detachment.
This inclusion and exclusion in plotting is also reflective to that historically done for BaSPMI and BaySPMI results.

Overall, the adaptive grid algorithm was able to correctly reproduce simulated ion sources and sinks within an uncertainty of one standard deviation.
For the most part, our analysis not only agrees with but provides greater certainty in ion sources and sinks than BaySPMI, reflected in the simulated inputs seen within narrower uncertainty bands.
However, this reduced quantile spread can be shown to give false certainty in specific cases, such as the large simulated spike of ionisation seen in the first case of Fig. \ref{fig:SimSourceSinkCompare}.

A noteworthy difference between the two analyses is that of ionisation inferred when the puffing rate is $2\times10^{21}$mol./s.
Full inclusion of simulated ionisation is seen in our adaptive grid approach whilst BaySPMI does not capture the simulation in the centre of the divertor chamber.
Since the Fulcher emissions were not behaving as expected, we set the maximum Fulcher emission we expect to see based upon a manual viewing of a map of Fulcher emission intensity within the divertor, resulting in different constraints for our $T_e^H$ parameter in the adaptive grid analysis.
This was not done for BaySPMI.

Whilst both analyses very slightly over-predict EIR near the target as gas puffing increases, essentially, the analyses indicates that EIR is negligible. 
This is coherent with early phases of detachment \cite{2022Spectroscopic}.

\subsubsection{Experimental Spectroscopic Data}

After having tested our analysis, we apply it to experimental data from discharge 46860 where the level of detachment is increased throughout the discharge as the core density is increased by additional core fuelling.

\begin{figure}[]
    \centering
    \includegraphics[width=0.95\textwidth]{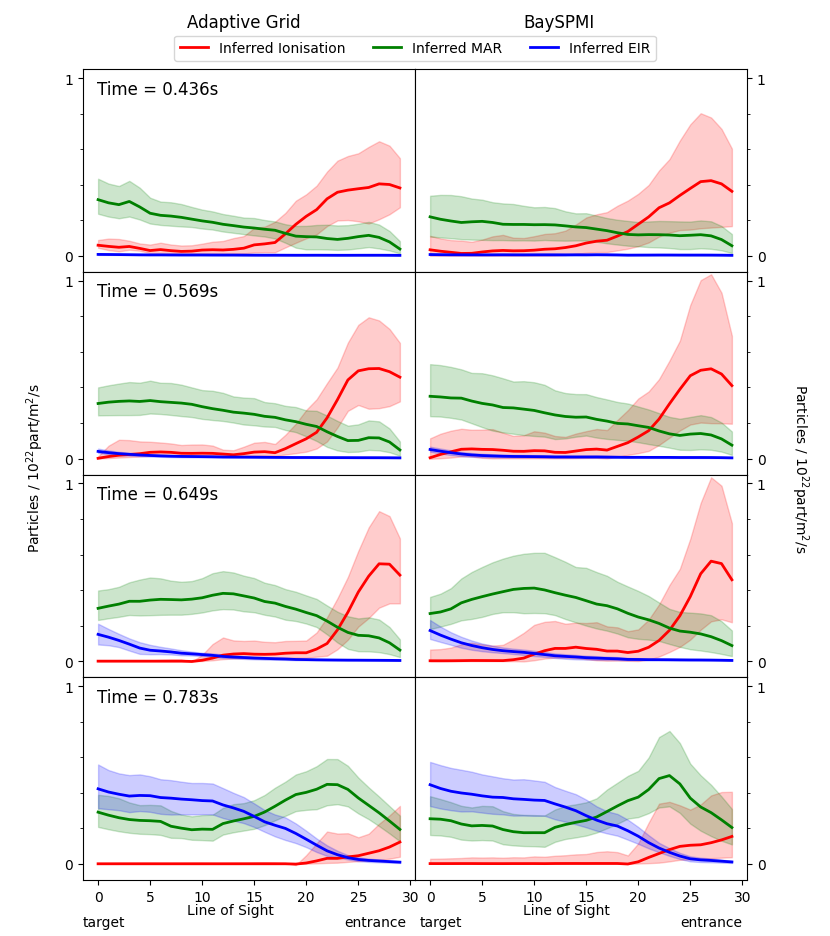}
    \caption{Ion and sink profiles from analysed results of measurements from discharge 46860 , using our adaptive grid analysis (left) and BaySPMI (right).
    Sources and sinks are given as a function of LoS and measured by number of particles undergoing these interaction at a given LoS and time.}
    \label{fig:RealSourceSinkCompare}
\end{figure}

We utilise conventional spectroscopy measurements of the $B_\alpha$, $B_\gamma$ and $B_\delta$ emissions as well as the $D_2$ Fulcher emission to construct a $T_e^E$ prior, manual adjustment of the Fulcher prior was not necessary.
Again, we run the analysis with all priors.
Model parameter comparisons were also made before post-processing for this analysis (App. C).
Comparative results for ion sources and sinks are given in Fig. \ref{fig:RealSourceSinkCompare}.

Very good agreement of inferred sources and sinks between the two different analyses is seen.
This is more impressive than that seen in Fig. \ref{fig:SimSourceSinkCompare} because both analyses display significantly thinner uncertainty bands whilst still agreeing.
The increased correspondence and narrower uncertainties are likely products of the inclusion of further Balmer lines resulting in better performance of the Bayesian analysis.
Most importantly, the adaptive grid analysis is able to infer the exact same stages of detachment, discussed further in Section \ref{sec:DetachmentResults}, seen in BaySPMI at the same points in time.

In the region where the ion source is insignificant, the adaptive grid shows smaller uncertainties than BaySPMI. 
However, both analyses indicate that the ionisation source is insignificant at these points.

\begin{figure}[]
    \centering
    \includegraphics[width=0.9\textwidth]{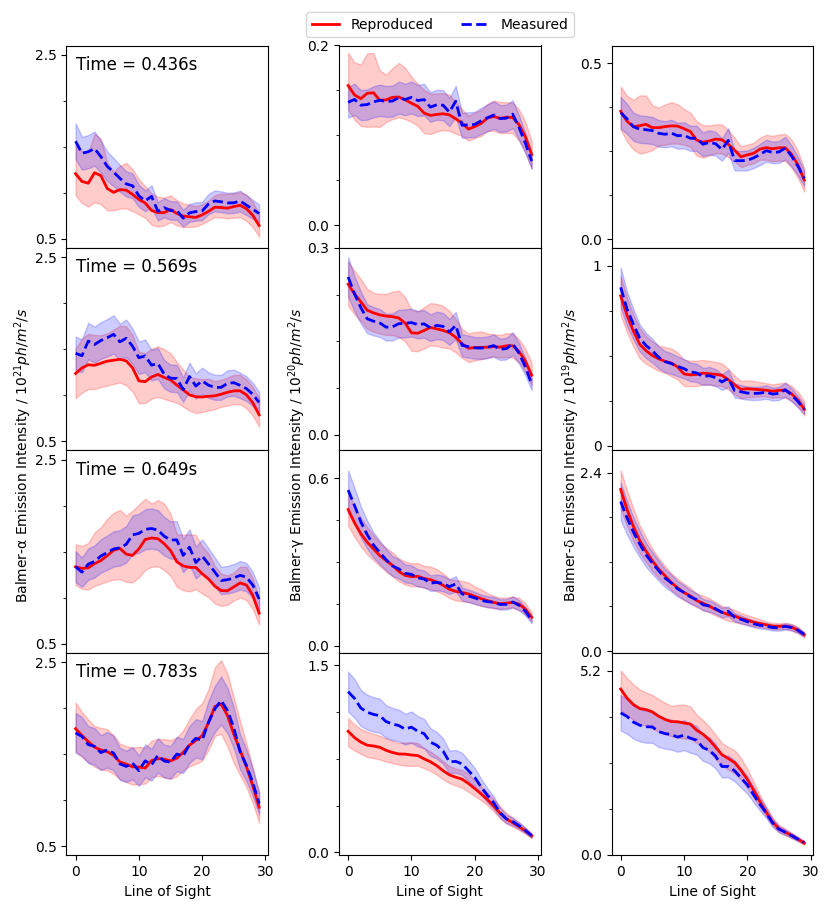}
    \caption{Balmer emission intensities as a function of LoS.
    Measured intensities (dashed-blue) are given with $\pm$12.5\%, reflective of equipment uncertainty.
    Reproduced intensities for the adaptive grid analysis of discharge 46860 were computed using a series of pre-developed post-processing tools, uncertainties are the 16 and 84\% quantiles, reflective of $\pm 1 \sigma$.}
    \label{fig:RealEmiss}
\end{figure}

Whilst important to verify our analysis against an already accepted method, this cannot be done if our analysis was to be used going forward as we would run into the same computational and time issues we are attempting to resolve.
Instead, we can reproduce the measured emission intensities, although this test cannot tell us about the whole validity of our results, it can tell us whether our model can correctly fit the data with our combinations of inferred parameters.

We move into a regime in which our experimental measurements significantly strays from our simplified emission model, thus the ability to reproduce observed intensities with such certainty (Fig. \ref{fig:RealEmiss}) is very impressive.
A very small gap between uncertainties is seen at time = 0.783s for $B_\gamma$, nevertheless, the majority of cases demonstrate that our analysis reproduces the emissions very well, giving confidence in both our model used and free parameters inferred.

\subsection{Accessibility of Software}

With the software meeting and exceeding expectations in validity, the accessibility of our approach is now considered.
We define a fair comparison between analysis methods as: when running the workstation at its limits, how fast does the analysis complete?
The results of this comparison are given in Tab. \ref{tab:Accessibility}.

\begin{table}[h]
\centering
\begin{tabular}{l|cccc}
                                                                  & \begin{tabular}[c]{@{}c@{}}Memory per\\ thread / GB\end{tabular} & Threads & \begin{tabular}[c]{@{}c@{}}Total\\ memory / GB\end{tabular} & Time    \\ \hline
\begin{tabular}[c]{@{}l@{}}Computational resources\end{tabular} &                                                                  & 64      & 128                                                           &         \\
BaSPMI                                                            & 3                                                              & 30$^{\#}$      & 90$^{\#\#}$                                                            & 48+ hr   \\
BaySPMI                                                           & 100+                                                             & 1$^*$       & 100+                                                          & 4-48 hr$^{**}$ \\
Adaptive Grid                                                     & 0.5                                                              & 60      & 30                                                            & < 3 mins 
\end{tabular}
\caption{Comparative computational resource use and time taken to run the numerous methods of analysis for detachment.
$^{\#}$BaSPMI is limited by the number of processes it can be split across, since each process must communicate with each other.
$^{\#\#}$Unaccounted for additional processes involved with BaSPMI analysis add to this memory usage. 
$^*$One thread can be used during the Bayesian analysis of the full grid, more can be used during sampling.
$^{**}$Dependent on divertor configuration.}
\label{tab:Accessibility}
\end{table}

The time taken for our adaptive script is that of discharge 46860 because this was a larger input (30 lines-of-sight across 33 time points, whilst the simulated data was 20 lines-of-sight across 35 time points).
Analysing the smaller-scaled simulated data took 2.5 minutes.
Where the input data shows divergence from MAST-U specific data, BaySPMI begins to find difficulty, hence the range in time values.
The 46860 discharge took 4 hours, whilst the simulated discharge took up to 48 hours.

An important result for our analysis is the considerably smaller memory usage than BaySPMI, due to significantly less parameter space being evaluated.
This ensures no risk of computational overload and the ability to run with high performance on a personal computer, unlike the adaptive grid approach in \cite{2019Bayesim}.
However, with access to the workstation used to run BaSPMI and BaySPMI, it allows the wide-scale multi-processing for our software to produce such fast results.

It should be noted that the time for BaSPMI includes the post-processing aspects of BaySPMI and our analysis.
However, to produce ion sources and sinks adds $\lesssim$ 0.5 minutes, and to reproduce emissions another $\lesssim$ 0.5 minutes.
Therefore, our analysis can easily complete within the 20 minutes between MAST-U pulses, allowing previously inaccessible insights into detachment processes before preceding discharges.

\subsection{Analysis of Detachment} \label{sec:DetachmentResults}

\begin{figure}[]
    \centering
    \includegraphics[width=0.9\textwidth]{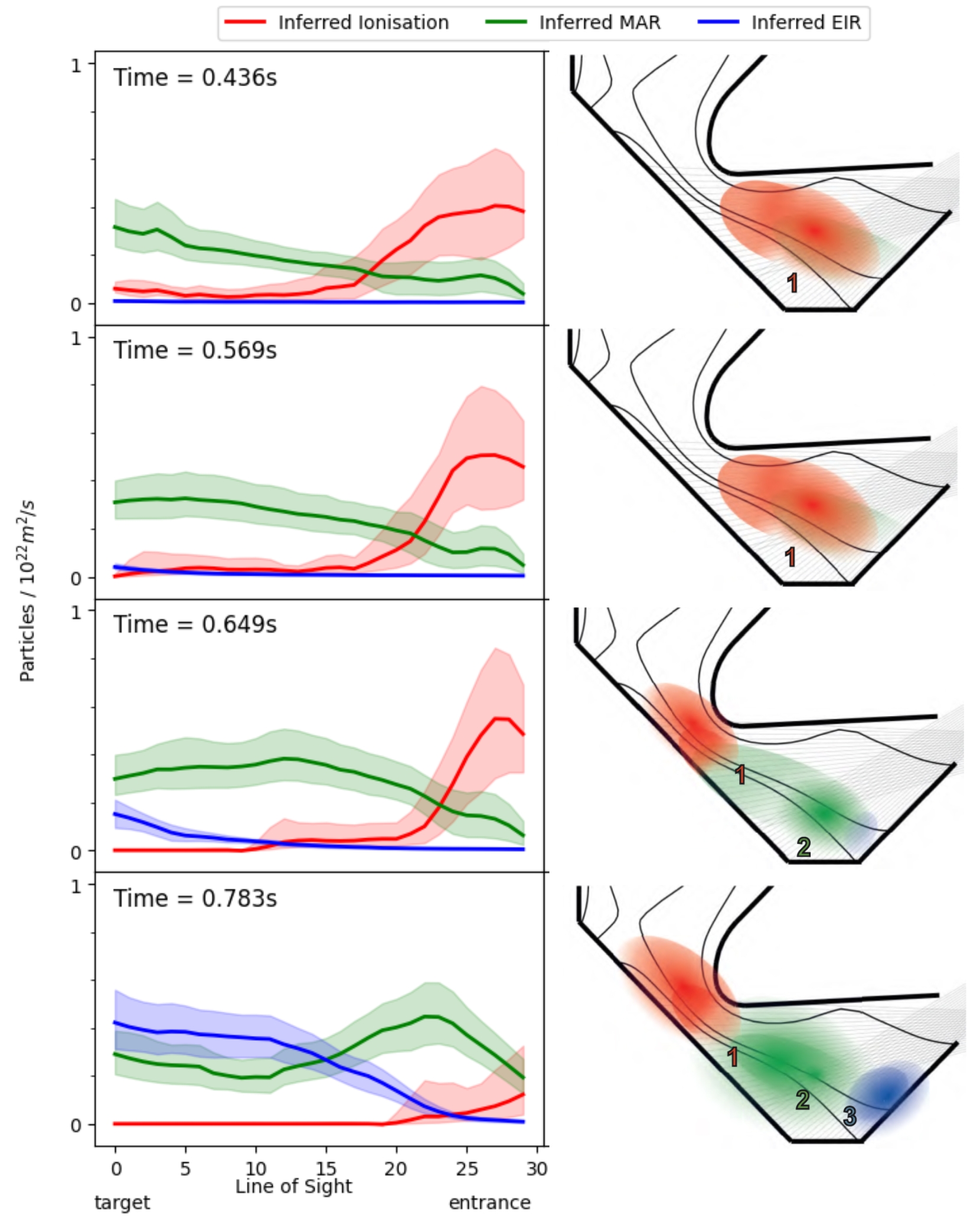}
    \caption{Ion sources and sinks, as inferred by our adaptive grid approach, for the 46860 MAST-U discharge, given as a function of LoS and measured in number of particles undergoing these interactions.
    Shown at four different points in time to demonstrate the time evolution for some of the phases of detachment.
    Schematic overview of detachment phases adopted from \cite{2022Spectroscopic}.}
    \label{fig:DetachmentAnalysis}
\end{figure}

First, we will discuss our observations during the SOLPS-ITER $N_2$ gas puffing experiment (Fig. \ref{fig:SimSourceSinkCompare}) \cite{myatra2021numerical, myatra2023predictive} in terms of the detachment physics involved. 
In this, a shift and detachment of the ionisation front is diagnosed as $N_2$ injection is increased and thus larger power losses occur (and thus more power limitation/ starvation, Eq. \ref{eq:FinalParticleBalance}), inferring the divertor is currently in phase one detachment \cite{2022Spectroscopic}. 
The comparative significance for ionisation against the ion sinks, could be explained by the higher temperatures in the divertor during $N_2$ gas puffing compared to a density ramp, which is not yet fully understood \cite{myatra2023predictive}. 
The lack of EIR and increase/movement of the MAR front is consistent with TCV experiments where no significant MAR or EIR is observed during $N_2$ seeding.

Moving from the $N_2$ seeded simulations to the MAST-U Super-X density ramp experiment 46860, excellent demonstration of the first three detachment phases is seen.
Our adaptive grid analysis is compared to the detachment phase infographic in Fig. \ref{fig:DetachmentAnalysis}.

Because the Super-X is so efficient at reducing heat flux and plasma temperatures in the divertor, discharge 46860 begins midway through phase 1 at time = 0.436s, as seen by the bulk of the ionisation being far from the target and a region of MAR already being established at the target.
The ion sources and sinks are shown at 0.569s to demonstrate the point of complete ionisation detachment from the target.

Our third time point in Fig. \ref{fig:DetachmentAnalysis} displays the plasma's movement into phase two.
Although the uncertainty on MAR does not yet infer a detachment of the front of the recombination region, our knowledge of detachment means that the onset of EIR at the target implies the shift in phase 2 and thus the movement of the MAR front.
Our last point in time demonstrates a clear detachment of the MAR front from the target and a large increase in EIR at the target.
This provides a clear implication that our system is now in phase three detachment.

If we explore the ion sources and sinks past 0.783s we do not see phase four detachment.
This is because further increasing the density would invite disruptions into the experimental set up of discharge 46860.
Lastly, it is implied that the ionisation front exits the divertor as identified in \cite{2022Spectroscopic}.
\section{Discussion}

\subsection{Results of Adaptive Grid Analysis}

Both full and adaptive grid Bayesian analyses demonstrate difficulty in inferring $f_\mathrm{mol}$ unless the actual value is near the parameter boundaries (i.e. majority atomic or molecular interactions).
This is because $f_\mathrm{mol}$ cannot be restricted by a prior probability due to minimal diagnostic measurements for the separation of molecular and atomic contributions to emitted Balmer lines.

However in terms of ion source/sink estimation and emission reproduction, the specifically uncertain $f_\mathrm{mol}$ causes no concern.
It is important to reiterate here that our analysis samples a six-dimensional parameter space, not six parameters individually.
Therefore, the varied results we infer for $f_\mathrm{mol}$ may correspond more specifically with certain combinations of our other free parameters.
Once these combinations are amalgamated in our emission model we are then given more certain outcomes.

The miss in ionisation peak in Fig. \ref{fig:SimSourceSinkCompare} at a puffing rate of $1\times10^{21}$mol./s is another result of interest, because this peak is included within the BaySPMI analysis.
A likely reason is the $T_e^E$ prior using a manually estimated maximum in Fulcher emission intensity.
It may be that this estimation requires further refining before better reproduction of simulated inputs.
Nevertheless, it would be ideal to avoid this manual effort of tweaking the analysis to allow smoother use of such a software.
Therefore, other methods of estimating a prior probability for $T_e^E$ would be welcomed into the script.
For example, knowledge that prolonged time during a density ramp experiment drives lower temperatures in the divertor could be used \cite{verhaegh2017spectroscopic}.

Even so, we can currently maintain confidence in our uncertainties of $\pm$1 standard deviation, without needing to expand to $\pm$2, due to the good agreement across other lines-of-sight and puff rates.

\subsection{Balmer Emission Model}

The Balmer emission model was originally simplified to six free parameters due to computational concerns for performing a full grid Bayesian analysis.
Our analysis demonstrates computational ease and efficiency for computing such a model over six-dimensional parameter space.
Therefore, there is potential to make this model more realistic with the introduction of further free parameters.

Although neglected throughout this report, negative molecular ions do play some role in the power and particle balance within the divertor \cite{verhaegh2021molecule, 2021Molecular}.
Hence, an additional ratio/fractional variable could be introduced within our $B_\mathrm{n \rightarrow 2}^\mathrm{mol}$ aspect of our emission model to consider the amount of contributions between negative and positive molecular ions.

Nevertheless, such improvements to the model may not ensure complete reproduction of emissions at high density (such as that seen in $B_\gamma$ at 0.783s in Fig. \ref{fig:RealEmiss}) due to the increased photon opacity associated with large densities \cite{terry1998opacity, pshenov2022opacity}.
For example, if there is a large population of neutral atoms surrounding our Balmer emitting interactions, the emitted photons could be absorbed and re-emitted at differing wavelengths, altering both the reaction rates found by our analysis as well as the PEC coefficients.

\subsection{Adaptive Grid Algorithm}

Our use of the adaptive grid algorithm has been specifically specialised for optimised use on MAST-U Super-X data, providing concerns that the current state of the software may not be applicable to other tokamaks yet.
However, the script was able to accurately reproduce input values based upon Super-X data, as well as ion sources/sinks within Super-X simulations.
Neither scenarios are completely identical to the real Super-X measurements, with the latter causing great difficulty for the BaySPMI software.
Therefore, our analysis shows good signs for adaptability for use in differing divertor systems.
Since our analysis is built upon BaSPMI which has been tested on TCV (using density ramp and $N_2$ seeded simulations and experiments) as well as DIII-D, there is a high probability that it would work.
Even so, testing on data from other tokamak systems is essential before we can have confidence in extended use of such an analysis.

The adaptive grid analysis shows outstanding increases in efficiency over its predecessors.
However it is yet to be tested on real ultrafast spectroscopic data, which can measure Balmer emission intensities at 800kHz across 10 lines-of-sight.
The maximum amount of time points from these 10 lines-of-sight that could be analysed by our script within 20 minutes is $\approx$700 (App D.).

Whilst such an analysis would be of better time resolution than that possible for conventional spectroscopic measurements, we cannot neglect the potential for further increases in efficiency.
This increased efficiency could be achieved in a couple of ways.
\begin{enumerate}
    \item Our analysis approach does not correctly converge for cell spacings that are of size equivalent to that of BaySPMI.
    This is because the algorithm only looks at the nearest neighbours directly adjacent to our current cell when initially climbing and filling the grid in probability.
    If cells are too large and we have complex six-dimensional shapes of probability, there is the possibility that the algorithm could miss areas of high probability that may have been `diagonal' to the current cell where finer cells would have gradually climbed to this area.
    Therefore, we could investigate a method of expanding the nearest neighbour search in order to utilise coarser cell spacing (equal to that of BaySPMI) for faster computation.
    \item The adaptive grid algorithm is currently set up in such a way that it must first evaluate a cell with a parameter outside of our hard limits before it knows this is an unrealistic hypotheses.
    It is able to identify that it must search no further around this cell after this evaluation.
    A method of acceptance/rejection of hypotheses based upon the parameter values before any evaluations in the first place could overcome these inefficiencies.
\end{enumerate}

It is worth noting that the ultrafast measurements are not intended for inter-shot analysis, but for specific analysis of shots after the fact, to study transient events.
Thus there are decreased time constraints for our analysis of ultrafast spectroscopic data.
\section{Conclusions}

We have presented the first application of a Bayesian inference technique for conventional and ultrafast spectroscopic measurements, utilising a new adaptive grid algorithm for probability analysis, to infer the physics within a divertor chamber during detachment.

A simplified emission model with six free parameters describing the Balmer line emissions due to plasma-atom (EIE and EIR) and plasma-molecular interactions was discussed and shown to successfully reproduce observed emission intensities.
The adaptive grid algorithm was established for use on observations from the Super-X divertor and used to infer the probabilities of these six free parameters, with minimal analysis inputs, which can be post-processed to obtain compelling estimates of ion sources/sinks and power loss mechanisms.
Methods for establishing the algorithm are accessible in the analysis script for ease of change across divertor configurations.

Initial testing within this work was performed on: ideal test cases to reproduce known parameter inputs; SOLPS-ITER simulations to reproduce known ion source/sink inputs; and MAST-U discharge 46860 to infer the same phases of detachment seen in BaySPMI analyses.
These tests have demonstrated the successful application for the analysis across all cases.

This analysis software also proves significant reductions in run-time and computational strain compared to previous analysis methods.
Thus, our foundational use of Bayesian techniques with adaptive grid algorithms for divertor diagnoses provides powerful understanding for plasma chemistry during detachment, previously unattainable between experimental discharges.
The flexibility of analyses associated with Bayesian approaches allows us to easily update our software for use with increased Balmer lines and prior conditions as they become available.

Future work will include further testing of this analysis software, across varying detachment scenarios and divertors, to build a more robust analysis script for wider-scale use.
Additionally, subsequent work on the efficiency of the adaptive script algorithm could lead to further applications across the tokamak diagnosis environment.

\section*{Acknowledgements}

This work utilises PEC databases made using ADAS \cite{o2013adas} and Yacora \cite{wunderlich2020yacora}, as well as simulations produced by Myatra, O, et al., \cite{myatra2023predictive} using SOLPS-ITER as the simulation software \cite{2015SOLPS}.
Thanks go to the whole diagnostic team at CCFE working on MAST-U, to produce experimental spectroscopic data for detachment.
Lastly, my thanks go to my supervisors at UKAEA, Verhaegh, K and Bowman, C, for providing continued knowledge and support throughout the completion of this project.

\printbibliography
\newpage 
\pagenumbering{Roman}
\subsection*{Appendix A. Analysis Script} \label{sec:Code}

\subsubsection*{Appendix A.1 User Script}

\begin{lstlisting}[language=Python]
from agsi.ReportAnalysis import FullAnalysis
import numpy as np

output = FullAnalysis('46860_BaySPMI_low_Te_150623_input.npy', specific_bound=True)

np.save('output_name.npy', output)

\end{lstlisting}

\subsubsection*{Appendix A.2 Analysis Script}

\begin{lstlisting}[language=Python]
from numpy import array, log, load, exp, linspace, ones, zeros, shape, nan, unravel_index, hstack
from pdfgrid import PdfGrid
from agsi.priors import *
from agsi.likelihood import *
from numpy.random import default_rng
import dms.analysis.emission.Balmer_analysis as BA
import matplotlib.pyplot as plt
import scipy.stats as scis
from multiprocessing import Pool
import numpy as np


def ImportData(input_data, band_check):
    Uncertainty = input_data['AbsErr']
    length_mode = input_data['DL']
    length_lower = input_data['DLL']
    length_upper = input_data['DLH']
    fulcher = input_data['Fulcher']
    fulcherLimits, te_lim_low, te_lim_high = BA.get_fulcher_constraint_spline(fulcher, telim=True, cummax=True)
    fulcherLimits = fulcherLimits
    DenMean = input_data['Den']
    DenErr = input_data['DenErr']
    n1 = int(input_data['n1'])
    n2 = int(input_data['n2'])

    if band_check == 3:
        D_alpha = input_data['n1Int']
        n2_line = input_data['n2Int']

        return D_alpha, n2_line, Uncertainty, length_mode, length_lower, length_upper, DenMean, DenErr, fulcherLimits, n1, n2

    else:
        D_alpha = input_data['DaMea']
        n1_line = input_data['n1Int']
        n2_line = input_data['n2Int']

        return D_alpha, n1_line, n2_line, Uncertainty, length_mode, length_lower, length_upper, DenMean, DenErr, fulcherLimits, n1, n2


def SpecificBounds(input_data):
    DenMax = input_data['DenMax']
    DenMin = input_data['DenMin']
    NeutralMax = input_data['noneH']
    NeutralMin = input_data['noneL']

    return DenMax, DenMin, NeutralMax, NeutralMin


def EmissionLines(n1, n2):
    n_lines = ["D_alpha", "D_beta", "D_gamma", "D_delta", "D_epsilon", "D_zeta", "D_eta"]
    if n1 == 3:
        line = n_lines[n1 - 3], n_lines[n2 - 3]
    else:
        line = n_lines[0], n_lines[n1 - 3], n_lines[n2 - 3]

    return line


def TwoBands(D_alpha, n2_line, uncertainty, length_mode, length_lower, length_upper, DenMean, DenErr, fulcherLimits, n1, n2, DenMax=4e19, DenMin=5e18, NeutralMax=1, NeutralMin=1e-3):

    if np.isnan(n2_line):
        adaptive_sample = np.full((500, 6), np.nan)

    else:
        lines = EmissionLines(n1, n2)
        simulated_data = array([D_alpha, n2_line])
        EmissionData = SpecData(
            lines=[lines[0], lines[1]],
            brightness=simulated_data,
            uncertainty=simulated_data*uncertainty
        )

        """
        Build posterior
        """
        likelihood = SpecAnalysisLikelihood(
            measurements=EmissionData
        )

        HardLimit = BoundaryPrior(
            upper_limits=array([log(5), log(1), log(DenMax), log(NeutralMax), 0.95, 0.4]),
            lower_limits=array([log(0.2), log(0.02), log(DenMin), log(NeutralMin), 0.01, 0.02])
        )

        ColdHotPrior = ColdTempPrior()

        length_prior = PathLengthPrior(mode=length_mode, lower_limit=length_lower, upper_limit=length_upper)

        FulcherPrior = HotTempPrior(fulcherLimits)

        DensityPrior = ElectronDensityPrior(mean=DenMean, sigma=DenErr)

        posterior = Posterior(
            components=[likelihood, ColdHotPrior, length_prior, FulcherPrior, DensityPrior],
            boundary_prior=HardLimit
        )

        """
        search for a good starting point
        """
        rng = default_rng()
        hypercube_samples = rng.uniform(low=HardLimit.lwr, high=HardLimit.upr, size=[100000, 6])
        hypercube_probs = posterior(hypercube_samples)
        grid_centre = hypercube_samples[hypercube_probs.argmax(), :]


        """
        Run the algorithm
        """
        grid_spacing = array([0.06, 0.09, 0.09, 0.15, 0.06, 0.02])
        grid_bounds = array([[log(0.2), log(0.02), log(DenMin), log(NeutralMin), 0.01, 0.02], [log(5), log(1), log(DenMax), log(NeutralMax), 0.95, 0.4]]).T
        grid = PdfGrid(spacing=grid_spacing, offset=grid_centre, bounds=grid_bounds, convergence=1e-1, n_samples=10000, n_climbs=200)

        while grid.state != "end":
            params = grid.get_parameters()
            P = posterior(params)
            grid.give_probabilities(P)

        adaptive_sample = grid.generate_sample(HardLimit, 500)

    return adaptive_sample


def ThreeBands(D_alpha, n1_line, n2_line, uncertainty, length_mode, length_lower, length_upper, DenMean, DenErr, fulcherLimits, n1, n2, DenMax=4e19, DenMin=5e18, NeutralMax=1, NeutralMin=1e-3):

    if np.isnan(n1_line) or np.isnan(n2_line):
        adaptive_sample = np.full((500, 6), np.nan)

    else:
        """
        Construct emission model
        """
        lines = EmissionLines(n1, n2)
        emission_data = array([D_alpha, n1_line, n2_line])
        EmissionData = SpecData(
            lines=[lines[0], lines[1], lines[2]],
            brightness=emission_data,
            uncertainty=emission_data*uncertainty
        )

        """
        Build posterior
        """
        likelihood = SpecAnalysisLikelihood(
            measurements=EmissionData
        )

        HardLimit = BoundaryPrior(
            upper_limits=array([log(5), log(1), log(DenMax), log(NeutralMax), 0.95, 0.4]),
            lower_limits=array([log(0.2), log(0.02), log(DenMin), log(NeutralMin), 0.01, 0.02])
        )

        ColdHotPrior = ColdTempPrior()

        length_prior = PathLengthPrior(mode=length_mode, lower_limit=length_lower, upper_limit=length_upper)

        FulcherPrior = HotTempPrior(fulcherLimits)

        DensityPrior = ElectronDensityPrior(mean=DenMean, sigma=DenErr)

        posterior = Posterior(
            components=[likelihood, ColdHotPrior, length_prior, FulcherPrior, DensityPrior],
            boundary_prior=HardLimit
        )

        """
        Run adaptive grid algorithm
        """
        grid_spacing = array([0.06, 0.09, 0.09, 0.15, 0.06, 0.02])
        grid_bounds = array([[log(0.2), log(0.02), log(DenMin), log(NeutralMin), 0.01, 0.02], [log(5), log(1), log(DenMax), log(NeutralMax), 0.95, 0.4]]).T
        grid_centre = (grid_bounds[:, 1] - grid_bounds[:, 0]) / 2
        grid = PdfGrid(spacing=grid_spacing, offset=grid_centre, bounds=grid_bounds, convergence=1e-1, n_samples=10000, n_climbs=200)

        while grid.state != "end":
            params = grid.get_parameters()
            P = posterior(params)
            grid.give_probabilities(P)

        adaptive_sample = grid.generate_sample(HardLimit, 500)

    return adaptive_sample


def OutputStructure(input, TeEMC, TeRMC, DenMC, noneMC, fmolMC, DLMC, PresMC):
    input['n1'] = int(input['n1'])
    input['n2'] = int(input['n2'])
    output = dict()
    output['input'] = input
    output['ResultMC'] = dict()
    output['ResultMC']['DenMC'] = DenMC
    output['ResultMC']['TeEMC'] = TeEMC
    output['ResultMC']['TeRMC'] = TeRMC
    output['ResultMC']['fmolMC'] = fmolMC
    output['ResultMC']['DLMC'] = DLMC
    output['ResultMC']['noneMC'] = noneMC
    output['ResultMC']['PresMC'] = PresMC

    return output


def FullAnalysis(input_file, poolcount=48, specific_bound=False):
    input_data = load(input_file, allow_pickle=True)
    input_data = input_data[()]
    band_check = input_data['n1']

    Iter = 500

    p = []
    pool = Pool(poolcount)

    if band_check == 3:
        D_alpha, n2_line, Uncertainty, length_mode, length_lower, length_upper, DenMean, DenErr, fulcherLimits, n1, n2 = ImportData(input_data, band_check)

        DenMC = zeros([shape(DenMean)[0], shape(DenMean)[1], Iter]) + nan
        TeEMC = zeros([shape(DenMean)[0], shape(DenMean)[1], Iter]) + nan
        TeRMC = zeros([shape(DenMean)[0], shape(DenMean)[1], Iter]) + nan
        DLMC = zeros([shape(DenMean)[0], shape(DenMean)[1], Iter]) + nan
        noneMC = zeros([shape(DenMean)[0], shape(DenMean)[1], Iter]) + nan
        fmolMC = zeros([shape(DenMean)[0], shape(DenMean)[1], Iter]) + nan
        PresMC = zeros([shape(DenMean)[0], shape(DenMean)[1], Iter]) + nan

        S = shape(DenMean)

        print('appending operations to pool')

        if specific_bound:
            DenMax, DenMin, NeutralMax, NeutralMin = SpecificBounds(input_data)

            for l in range(0, S[0]*S[1]):
                i, j = unravel_index(l, S)

                p.append(pool.apply_async(TwoBands, (D_alpha[i, j], n2_line[i, j], Uncertainty, length_mode[i, j], length_lower[i, j], length_upper[i, j], DenMean[i, j], DenErr[i, j], fulcherLimits[i, j], n1, n2), dict(DenMax=DenMax, DenMin=DenMin, NeutralMax=NeutralMax[i, j], NeutralMin=NeutralMin[i, j])))

        else:
            for l in range(0, S[0] * S[1]):
                i, j = unravel_index(l, S)

                p.append(pool.apply_async(TwoBands, (D_alpha[i, j], n2_line[i, j], Uncertainty, length_mode[i, j], length_lower[i, j], length_upper[i, j], DenMean[i, j], DenErr[i, j], fulcherLimits[i, j], n1, n2)))

        print('executing pool')
        sample = [p[hh].get() for hh in range(len(p))]

        for l in range(0, S[0]*S[1]):
            i, j = unravel_index(l, S)

            TeEMC[i, j, :] = exp(sample[l][:, 0])
            TeRMC[i, j, :] = exp(sample[l][:, 1])
            DenMC[i, j, :] = exp(sample[l][:, 2])
            noneMC[i, j, :] = exp(sample[l][:, 3])
            fmolMC[i, j, :] = sample[l][:, 4]
            DLMC[i, j, :] = sample[l][:, 5]
            PresMC[i, j, :] = exp(sample[l][:, 0]+sample[l][:,2])

        output = OutputStructure(input_data, TeEMC, TeRMC, DenMC, noneMC, fmolMC, DLMC, PresMC)

    else:
        D_alpha, n1_line, n2_line, Uncertainty, length_mode, length_lower, length_upper, DenMean, DenErr, fulcherLimits, n1, n2 = ImportData(input_data, band_check)

        DenMC = zeros([shape(DenMean)[0], shape(DenMean)[1], Iter]) + nan
        TeEMC = zeros([shape(DenMean)[0], shape(DenMean)[1], Iter]) + nan
        TeRMC = zeros([shape(DenMean)[0], shape(DenMean)[1], Iter]) + nan
        DLMC = zeros([shape(DenMean)[0], shape(DenMean)[1], Iter]) + nan
        noneMC = zeros([shape(DenMean)[0], shape(DenMean)[1], Iter]) + nan
        fmolMC = zeros([shape(DenMean)[0], shape(DenMean)[1], Iter]) + nan
        PresMC = zeros([shape(DenMean)[0], shape(DenMean)[1], Iter]) + nan

        S = shape(DenMean)

        print('appending operations to pool')

        if specific_bound:
            DenMax, DenMin, NeutralMax, NeutralMin = SpecificBounds(input_data)

            for l in range(0, S[0] * S[1]):
                i, j = unravel_index(l, S)

                p.append(pool.apply_async(ThreeBands, (D_alpha[i, j], n1_line[i, j], n2_line[i, j], Uncertainty, length_mode[i, j], length_lower[i, j], length_upper[i, j], DenMean[i, j], DenErr[i, j], fulcherLimits[i, j], n1, n2), dict(DenMax=DenMax, DenMin=DenMin, NeutralMax=NeutralMax[i, j], NeutralMin=NeutralMin[i, j])))

        else:
            for l in range(0, S[0] * S[1]):
                i, j = unravel_index(l, S)

                p.append(pool.apply_async(ThreeBands, (D_alpha[i, j], n1_line[i, j], n2_line[i, j], Uncertainty, length_mode[i, j], length_lower[i, j], length_upper[i, j], DenMean[i, j], DenErr[i, j], fulcherLimits[i, j], n1, n2)))

        print('executing pool')
        sample = [p[hh].get() for hh in range(len(p))]

        for l in range(0, S[0] * S[1]):
            i, j = unravel_index(l, S)

            TeEMC[i, j, :] = exp(sample[l][:, 0])
            TeRMC[i, j, :] = exp(sample[l][:, 1])
            DenMC[i, j, :] = exp(sample[l][:, 2])
            noneMC[i, j, :] = exp(sample[l][:, 3])
            fmolMC[i, j, :] = sample[l][:, 4]
            DLMC[i, j, :] = sample[l][:, 5]
            PresMC[i, j, :] = exp(sample[l][:, 0]+sample[l][:, 2])

        output = OutputStructure(input_data, TeEMC, TeRMC, DenMC, noneMC, fmolMC, DLMC, PresMC)

    return output
\end{lstlisting}

\clearpage

\subsection*{Appendix B. Ideal Test Cases}

\subsubsection*{Appendix B.1 Ideal Test Case Set-Up}

Our ideal test case scenario was produced by feeding a realistic combination of parameters (Tab. \ref{tab:SyntheticParams}) into our emission model (Eq. \ref{eq:EmissionModel}) to produce intensities of Balmer lines for use in our analysis.

\begin{table}[h]
\centering
\begin{tabular}{l|cccccc}
                                                           & $T_e^E$ / eV & $T_e^R$ / eV & $n_e$ / $\mathrm{m^{-3}}$ & $n_\mathrm{ratio}$ & $f_\mathrm{mol}$ & $\Delta L$ / m\\ \hline
\begin{tabular}[c]{@{}l@{}}Test \\ Value\end{tabular} & 4.0   & 0.8   & 1.7 $\times 10^{19}$ & 0.15      & 0.35    & 0.09    
\end{tabular}
\caption{Input parameter values used to generate Balmer emission lines that can be analysed by our software.
To best ensure a practical combination of free parameters, these values were influenced by the resultant parameters from the BaySPMI analysis of pulse 46860.}
\label{tab:SyntheticParams}
\end{table}

A comparative full grid approach for the Bayesian analysis used in the adaptive grid script was produced to provide an additional test of validity.
Such a comparison can verify whether any failed reproductions of the ideal test data is due to the adaptive grid approach, or the Bayesian analysis and subsequent emission model itself.
The full grid script uses 25 equally spaced values within the limits of each parameter to produce the parameter space. 

Prior probability incorporation within the ideal test case has been reduced to prevent the potential for over-complication leading to unrealistic test scenarios.
Therefore, the utilisation of Fulcher emissions and Stark broadening for temperature and electron density priors respectively, have been neglected.
Hard limits for the parameters are those stated in Section \ref{sec:PriorImplication}.
Prior models for the comparative cold/hot temperatures and path length are kept, due to certainty of $T_e^R$ < $T_e^E$ and strong knowledge of the path length parameter in MAST-U.

\subsubsection*{Appendix B.2 Ideal Test Case Emission Reproduction}

\begin{figure}
    \centering
    \includegraphics[width=0.85\textwidth]{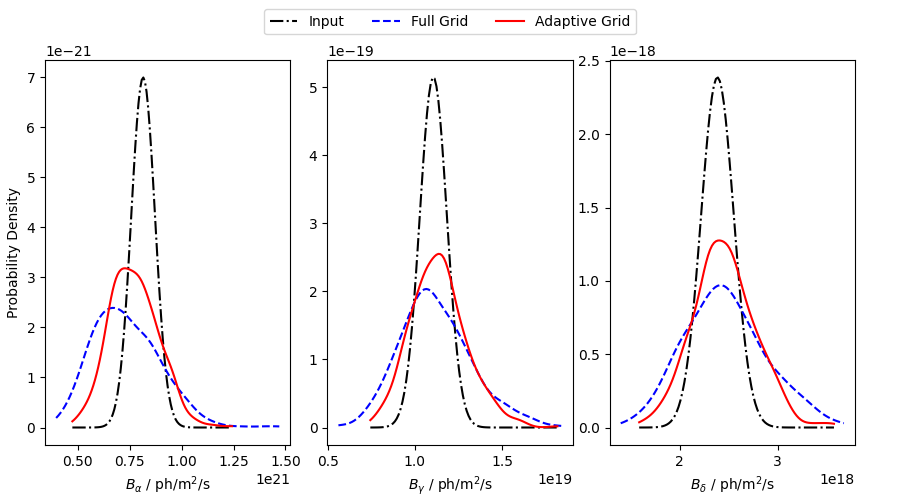}
    \caption{PDFs of Balmer-$\alpha$, $\gamma$ and $\delta$ line emission intensities reproduced by adaptive (straight-red) and full (dashed-blue) analyses.
    The inputted model emission (dashed-dotted) was plotted as a Gaussian with a standard deviation of 0.125, reflecting our 12.5\% equipment uncertainty.
    Emissions were reproduced by sampling real parameter values from our six-dimensional PDF and feeding these samples back into our model.
    Kernel density estimation \cite{chen2017kernel} was utilised to convert samples into PDFs.
    Probabilities were normalised via division by integral of PDF.}
    \label{fig:ThreeBandSynEmis}
\end{figure}

We plot the PDFs of our reproduced emissions for the contributing Balmer lines in Fig. \ref{fig:ThreeBandSynEmis}.
Although our ideal test case utilises the same model to create the emissions, such a plot can tell us whether our sampled output parameters can correctly fit the data.
Considering the consistency between parameter input and PDF in Fig. \ref{fig:ThreeBandSynParams}, good agreement is also seen between reproduced and model emissions across all Balmer lines.
Confidence for our adaptive analysis is given by the reproduction of all possible model emission inputs.
Whilst we do reproduce emissions outside of our inputs, these are of low probability density, and are also seen within the full grid approach.
Lastly, the bulk of our reproduced PDFs remain within the model Gaussian for Balmer emission intensities.

\begin{figure}
    \centering
    \includegraphics[width=0.9\textwidth]{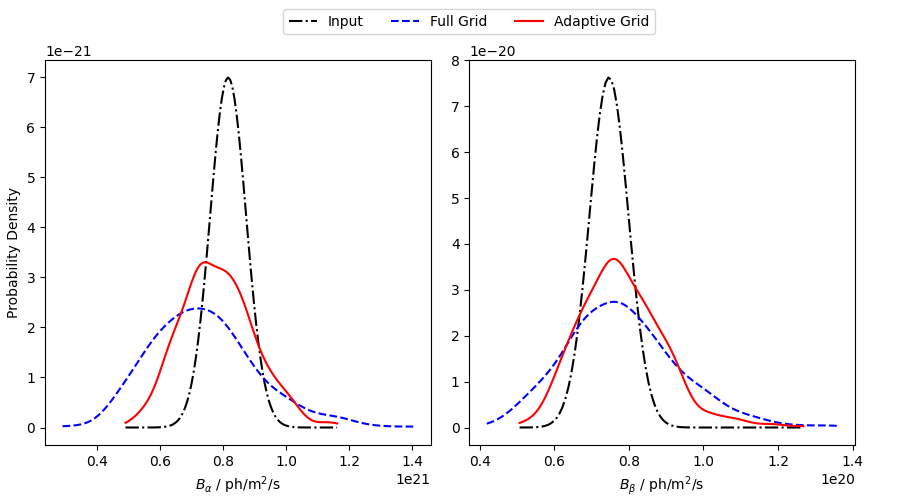}
    \caption{PDFs of Balmer-$\alpha$ and $\beta$ line emission intensities reproduced by adaptive (straight-red) and full (dashed-blue) analyses, with model emissions shown in dashed-dotted lines.
    Probabilities were normalised via division by integral of PDF.
    All PDFs were produced as in Fig. \ref{fig:ThreeBandSynEmis}.}
    \label{fig:TwoBandSynEmis}
\end{figure}

The same plot was made for the ultrafast spectrometer ideal test case, again, all model emission inputs are reproduced by our analysis (Fig. \ref{fig:TwoBandSynEmis}), inferring that the incorrect peak in $f_\mathrm{mol}$ in Fig. \ref{fig:TwoBandSynParams} is not detrimental to the model in full.

\clearpage

\subsection*{Appendix C. Free Parameter Validation} \label{sec:Params}

\begin{figure}[h]
    \centering
    \includegraphics[width=0.9\textwidth]{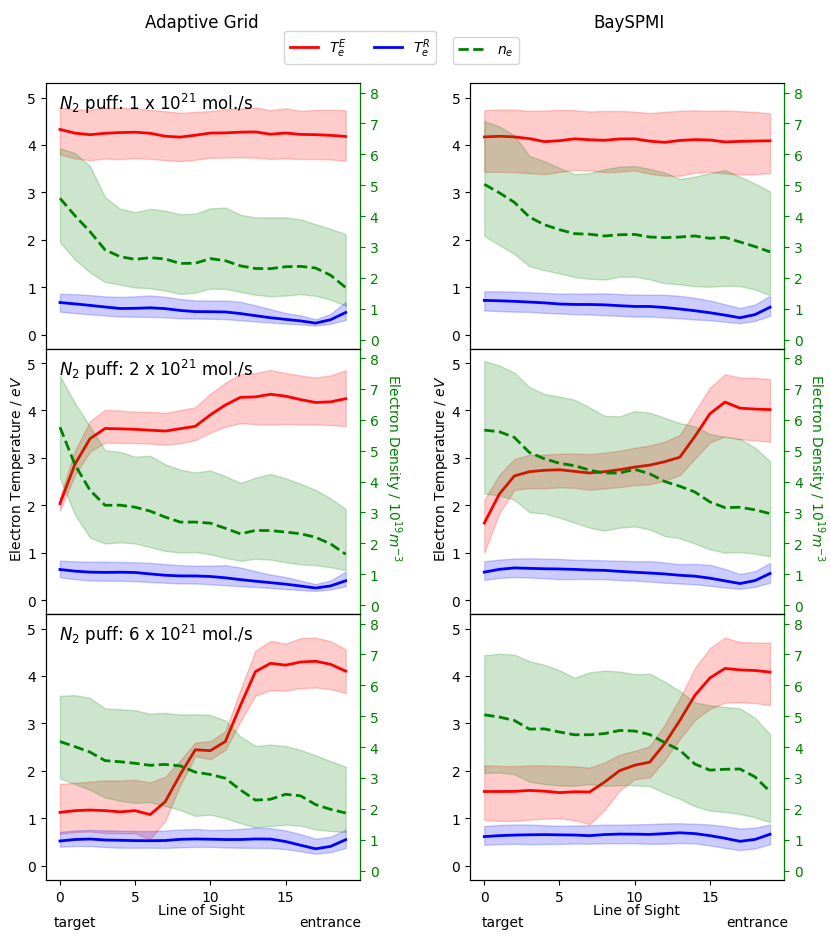}
    \caption{Important free parameters, inferred from measurements from an N2 gas puffing simulation in a Super-X-like divertor system using our adaptive grid analysis (left) and BaySPMI (right).
    Parameters are given as a function of LoS and measured by number of particles undergoing these interaction at a given LoS and time.}
    \label{fig:SimParamsCompare}
\end{figure}

\begin{figure}
    \centering
    \includegraphics[width=0.9\textwidth]{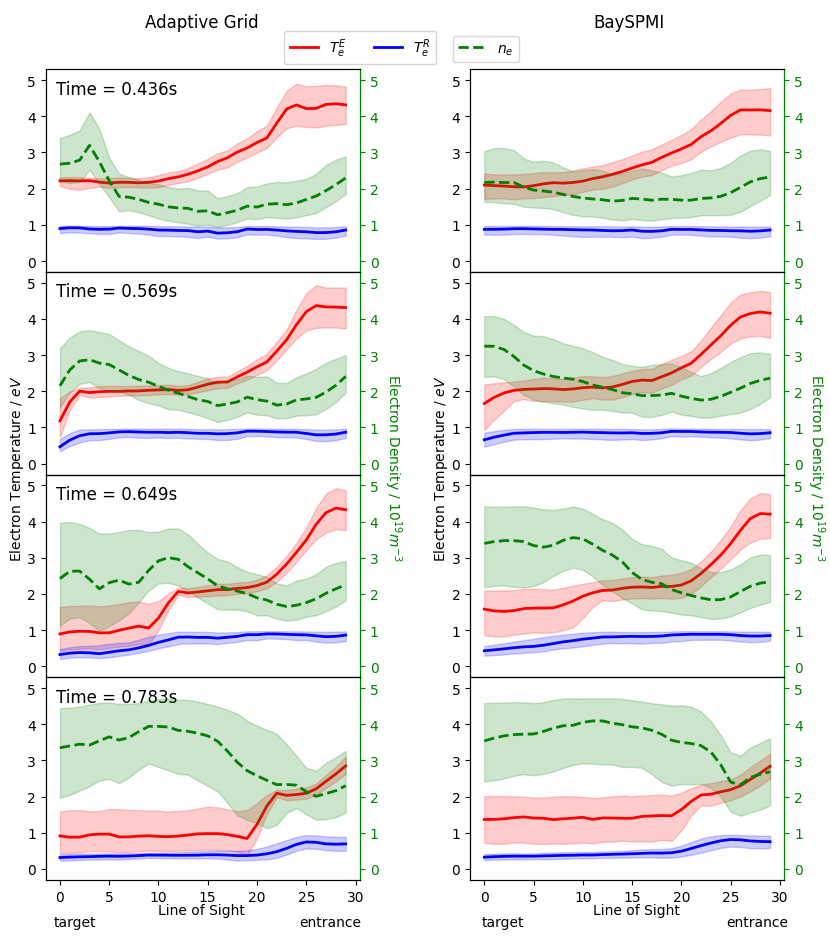}
    \caption{Important free parameters, inferred from experimental observations of MAST-U discharge 46860 using our adaptive grid analysis (left) and BaySPMI (right).
    Parameters are given as a function of LoS and measured by number of particles undergoing these interaction at a given LoS and time.}
    \label{fig:RealParamCompare}
\end{figure}

Before post-processing was performed on the sample parameter results from the adaptive grid and BaySPMI analyses, important parameters were compared for both simulated ultrafast (Fig. \ref{fig:SimParamsCompare}) and experimental conventional spectroscopic inputs (Fig. \ref{fig:RealParamCompare}).
Agreement via overlapping uncertainties between analyses can be seen across all parameters for all times and LoS, giving confidence in the parameter inferrence of our analysis.

\clearpage

\subsection*{Appendix D. Maximum Frequency Analysis}

Our analysis script is shown to complete investigation of Balmer emission intensities for 30 lines-of-sight across 33 points in time within 180 seconds.
15 of these 180 seconds account for setting up the analysis.
Thus, our time for evaluations is 165 seconds, and our time per evaluation (t) is $\approx$0.17s.

To analyse ultrafast spectroscopic measurements across 10 lines-of-sight within 20 minutes (1200 seconds) we define the following equation, in which f is frequency of measurements.

\begin{equation}
\begin{split}
    f \times LoS \times t &= 1200\\
    f &= 700
\end{split}
\end{equation}

Thus we can only analyse 700 time points for ultrafast spectroscopic measurements within 20 minutes.


\end{document}